\documentclass[manuscript,screen]{acmart}
\AtBeginDocument{%
  }

\setcopyright{acmlicensed}
\copyrightyear{2025}
\acmYear{2025}
\acmDOI{XXXXXXX.XXXXXXX}
\acmConference[Conference acronym 'XX]{Make sure to enter the correct
  conference title from your rights confirmation email}{June 03--05,
  2018}{Woodstock, NY}
\acmISBN{978-1-4503-XXXX-X/2018/06}

\usepackage{graphicx}
\usepackage{xspace}
\usepackage{enumitem}
\usepackage{amsmath,amsfonts}
\usepackage{algorithmic}
\usepackage{graphicx}
\usepackage{textcomp}
\usepackage{makecell}
\usepackage{balance}
\usepackage{caption}
\usepackage{hyperref}
\usepackage{seqsplit}
\usepackage[normalem]{ulem}
\begin{document}

\title{Understanding Inconsistent State Update Vulnerabilities in Smart Contracts}

\author{Lantian Li}
\affiliation{%
  \institution{Shandong University}
  \city{Qingdao}
  \country{China}
}
\email{lilantian@mail.sdu.edu.cn}

\author{Yuyu Chen}
\affiliation{%
  \institution{Shandong University}
  \city{Qingdao}
  \country{China}
}
\email{yuyuchen@mail.sdu.edu.cn}

\author{Jingwen Wu}
\affiliation{%
  \institution{Shandong University}
  \city{Qingdao}
  \country{China}
}
\email{elowen.jjw@gmail.com}

\author{Yue Pan}
\affiliation{%
  \institution{Shandong University}
  \city{Qingdao}
  \country{China}
}
\email{pany@mail.sdu.edu.cn}

\author{Zhongxing Yu}
\authornote{Zhongxing Yu is the corresponding author.}
\affiliation{%
  \institution{Shandong University}
  \city{Qingdao}
  \country{China}
}
\email{zhongxing.yu@sdu.edu.cn}

\renewcommand{\shortauthors}{Li et al.}

\begin{abstract}
Smart contracts enable contract terms to be automatically executed and verified on the blockchain, and recent years have witnessed numerous applications of them in areas such as financial institutions and supply chains. The execution logic of a smart contract is closely related to the contract state, and thus the correct and safe execution of the contract depends heavily on the precise control and update of the contract state. However, the contract state update process can have issues. In particular, inconsistent state update issues can arise for reasons such as unsynchronized modifications. Inconsistent state update bugs have been exploited by attackers many times, but existing detection tools still have difficulty in effectively identifying them. This paper conducts the first empirical study about inconsistent state update vulnerabilities (\emph{i.e.,} inconsistent state update bugs that are exploitable) in smart contracts, aiming to shed light for developers, researchers, tool builders, and language or library designers in order to avoid inconsistent state update vulnerabilities. We systematically investigate 116 inconsistent state update vulnerabilities in 352 real-world smart contract projects, summarizing their root causes, fix strategies, and exploitation methods. Our study provides {11} original and important findings, and we also give the implications of our findings. To illustrate the potential benefits of our research, we also develop a proof-of-concept checker based on one of our findings. The checker effectively detects issues in 64 popular GitHub projects, and 19 project owners have confirmed the detected issues at the time of writing. The result demonstrates the usefulness and importance of our findings for avoiding inconsistent state update vulnerabilities in smart contracts.
\end{abstract}

\ccsdesc[500]{Software and its engineering~Maintaining software}
\ccsdesc[500]{Software and its engineering~Language features}

\keywords{Smart contract, Inconsistent state update, Software security}


\maketitle

\section{Introduction}
\label{intro}
Ethereum is widely recognized as a globally decentralized computing infrastructure, and it uses the blockchain to synchronize and store the state changes of the system \cite{mastering}. In particular, the Ethereum platform can run programs called \emph{smart contracts}, which enable contract terms to be automatically executed and verified on the blockchain. Due to the convenience offered by high-level programming languages (\emph{e.g.}, Solidity) for writing contracts and the security guarantees provided by the underlying consensus protocol, recent years have witnessed various applications of smart contracts in areas such as financial institutions \cite{Financial}, online gaming \cite{gaming}, government governance \cite{government}, and supply chains \cite{electronics12061340}. Consequently, smart contract vulnerabilities can result in significant losses, as evidenced by recent attacks on smart contracts \cite{dao, Wallet, survey}. As an example, the infamous ``TheDAO'' reentrancy attack caused the loss of over \$50 million worth of Ethers. Worse still, smart contracts are known for their immutable nature--once deployed on the blockchain, it is tremendously difficult to fix bugs. While proxy contracts allow updates to the contract’s logic, modifying a deployed contract comes with considerable cost and risk due to the need to manage data migration, state consistency, and logic separation \cite{bodell2023proxy}.
In view of this aspect, vetting the contract's quality before deployment is even more important than that for traditional software.

The execution logic of a smart contract on the Ethereum platform is closely related to the contract state, and thus the correct and safe execution of the contract depends heavily on the precise control and update of the contract states. For smart contracts, 
state variables (also called storage variables) store the persistent data of the contracts and the state variables in turn are recorded in the blockchain, constituting the contract state \cite{bose2022sailfish}. Smart contract state variables are often implicitly correlated such that modifications to one variable typically necessitate synchronized updates of other related variables \cite{torres2021confuzzius}. However, developers are not always clear enough about this synchronization \cite{dedaubblog,quicknodeguides}. When updating a state variable, they may forget to update the correlated state variables or incorrectly update the correlated state variables, leading to inconsistent state update bugs. 

Inconsistent state update bugs are common in practice, and a recent study shows that they account for 18\% of the total Code4rena bugs \cite{zhang2023demystifying}. Thus, proper identification of correlations between state variables and achieving synchronized updates constitute a fundamental task in smart contract development. 
Depending on the specific state variables involved, the consequences of inconsistent state update bugs rang from wrong statistics to financial losses. In fact, several well-known smart contract attacks (such as the DAO attack, the Parity wallet hack, \emph{etc.}) are closely related to inconsistent state update bugs \cite{brent2020ethainter, perez2021smart}. Specifically, the DAO attack falls under the category of Interim State Exploit in the classification of exploitation methods presented in Sec.~\ref{secExploitation} of this paper.
Worse still, vulnerability attacks using inconsistent state update bugs are on the rise and account for 11\% of the real-world exploits \cite{zhang2023demystifying}. For instance, three recent exploits have resulted in around \$3.8 million loss \cite{exploit1,exploit2,exploit3}. 
Given this, it is of vital importance for avoiding inconsistent state update bugs that are exploitable (referred to as inconsistent state update vulnerabilities for simplicity). 

However, while much progress has been made in smart contract security research \cite{ren2021empirical, sendner2024large, wu2024we}, it is extremely difficult to effectively audit inconsistent state update vulnerabilities with existing automated tools \cite{zhang2023demystifying}. 
The main reason is that existing tools only check limited properties (\emph{e.g.}, detect access
violations through \texttt{tx.origin}), however inconsistent state update vulnerabilities involve complex correlations between state variables. Thus, inconsistent state update vulnerabilities remain a significant issue to the security of smart contracts and dedicated tools and techniques are in great demand. 

\textbf{Perspective and Goal.} We believe that to ameliorate this issue, a good understanding of inconsistent state update vulnerabilities is a prerequisite. In light of this, we in this paper conduct the first empirical study on inconsistent state update vulnerabilities in the wild. In particular, our goal is to shed light on these {three} research questions:

\begin{itemize}[]
\item {\textbf{RQ1 (Root causes for inconsistent state update vulnerabilities): What are the root causes of such vulnerabilities?} This question aims to help understand which factors may lead to inconsistent state updates during the development process, and then provide guidance for vulnerability detection.}

\item {\textbf{RQ2 (Fix strategies for inconsistent state update vulnerabilities): How can developers fix such vulnerabilities to avoid causing harm?} This question aims to explore effective fix strategies and best practices to help development teams take appropriate measures after discovering inconsistent state update vulnerabilities.

\item {\textbf{RQ3 (Exploitation methods for inconsistent state update vulnerabilities): In what forms are inconsistent state update vulnerabilities exploited?} This question aims to understand the technical means and attack patterns of attackers to exploit such vulnerabilities, and then provide clues for helping developers prevent similar attacks.}}
\end{itemize}

\textbf{Method and Subjects.} To answer these {three} questions, we studied 116 inconsistent state update vulnerabilities in 352 real-world smart {contract projects} from 2021 to 2024. These vulnerability samples come from the prestigious Code4rena competition \cite{Code4rena}, which invites individuals and companies around the world to participate and incentivizes participants to audit real-world smart contracts by offering generous bonuses \cite{Leaderboard}. 
For the contract projects to which these 116 vulnerabilities belong,
the average LLOC (logical lines of code, without comments and white spaces) \cite{ajienka2020empirical} is 14603.51, the average NF (number of functions) is 1703.14, and the average number of state variables is 683.67.
For each of the 116 inconsistent state update vulnerabilities, we deeply analyze the source code (which is available through GitHub), code comments, bug reports, and fix suggestions of the relevant contracts to fully understand the root causes, effective fix strategies, and  exploitation methods. 
To illustrate the potential benefits of our research, we developed a proof-of-concept checker to detect state variable update omissions and state variable optimization issues by checking whether there are state variables that are not declared with \texttt{constant} or \texttt{immutable} modifier and have never been reassigned. 
For evaluating the effectiveness of the checker,
we selected 208 active and popular projects from GitHub based on the criteria that they have been updated within the past month (relative to our evaluation time) and have at least 30 stars, and used them as samples for the checker evaluation.

{\textbf{Results.} Based on the study results of root causes, fix strategies, and exploitation methods, we systematically classify the 116 inconsistent state update vulnerabilities into different categories from these three dimensions. For each category, we quantify its distribution and summarize its pattern. To enhance the understanding, we illustrate through numerous bug examples during the process. 
Meanwhile, we also present the correlations between root causes and their corresponding fix strategies as well as exploitation methods. Overall, our study enables us to deliver 11 original and important findings. We here highlight several findings: 
\begin{itemize}[]
\item {The most significant root cause (47.4\%) is that for transactions involving multiple operation steps, some state variables fail to be updated in time after each operation. In particular, this category of bugs occurs frequently for state synchronization in permission management.}

\item {The second most significant root cause (34.48\%) is incorrect logic underlying the update computation, particularly in terms of improper call sequence and improper boundary condition handling.}

\item {The root causes can also be absence of expected critical state variables and the absence of  explicit initializations or re-initializations of certain critical state variables.}

\item {The most significant fix strategy (58.62\%) is directly modifying computations on the unsafe state variables, in particular time-related state variables.}

\item {The most significant exploitation method (56.03\%) is making use of numerical calculation errors, meaning that attackers seek and exploit computational discrepancies to amplify their impact and gain unfair advantages.}

\item {The second most significant exploitation method (23.28\%) is using repeated transactions, where an attacker triggers operations multiple times within a short period and exploits stale state data to obtain repeated profits.}

\item {The fix strategies are highly correlated with the causes, but the correlation between causes and exploitation methods
is relatively weak.}

\end{itemize}

We also give the implications of our findings, which shed light for developers, researchers, tool builders, and language or library designers in order to improve all facets of state variable updates in smart contracts. 
Furthermore, we present the evaluation results of our proof-of-concept checker, which was applied to currently active and popular Solidity projects. The checker detected issues in the latest code versions of 64 projects, and 19 project owners have confirmed the detected issues. Our checker has received positive feedback from many Solidity developers, with some describing our research as ``meaningful'' and others calling it a ``good shout!''. These results show that our findings can indeed help developers avoid inconsistent state update vulnerabilities to a certain extent during the development process, and also provide effective reference for builders of smart contract security detection tools.

\textbf{Contributions.} This paper makes the following main contributions:

\begin{itemize}[]
\item {We conduct the first systematic empirical study of inconsistent state update vulnerabilities in the wild.}

\item {We classify inconsistent state update vulnerabilities into different categories based on their root causes, fix strategies, and exploitation methods, and analyze each category to obtain 11 original and important findings with actionable implications. These findings and implications are beneficial to developers, researchers,  tool builders, and language or library designers.}

\item {We develop a checker to detect missed state variable updates or extra gas consumption by checking whether there are state variables that are not declared with \texttt{constant} or \texttt{immutable} modifier and are not reassigned throughout the project, and the usefulness of the checker has been confirmed by developers.}
\end{itemize}

\textbf{Paper Structure.} The remainder of this paper is structured as follows. We
first present necessary background in Sec.~\ref{Background}. Sec.~\ref{Methodology} explains the methodology to perform the empirical study. Sec.~\ref{causes}, Sec.~\ref{fix}, and Sec.~\ref{exploit}
present the results of the empirical study about the root causes, fix strategies, and exploitation methods for inconsistent state update vulnerabilities respectively, followed by Sec.~\ref{checker}
which presents the designed proof-of-concept checker. The last two sections give some closely related work and the conclusion respectively.

\section{Background}
\label{Background}
In this section, we give the necessary background on smart contracts, state variables and modifiers, and inconsistent state update bugs.
\subsection{Smart Contracts}
Ethereum smart contracts are essentially Turing-complete programs, and are compiled down to EVM (Ethereum Virtual Machine) bytecode before execution on the blockchain \cite{nguyen2020sfuzz}. Smart contracts can be written in specifically designed high-level languages such as Solidity, Vyper, and Serpent. Among these languages, Solidity is the most popular and widely considered by the community as the de facto language \cite{mastering,liIDOL,Solsmith}. Solidity is categorized as an \emph{imperative} language, and its syntax is similar to that of popular imperative languages like Java, C++, \emph{etc.} In particular, smart contracts also consist of a set of functions and variables. 
Smart contracts enable contract terms to be automatically executed and verified on the blockchain, and the verification especially is governed by the underlying consensus protocol. For both the convenience offered by high-level programming languages and the security guarantees provided by consensus protocol, smart contracts are widely used in areas such as financial institutions \cite{Financial}, online gaming \cite{gaming}, government governance \cite{government}, and supply chains \cite{electronics12061340}.

\subsection{State Variables and Modifiers}
Ethereum smart contracts use state variables (also called storage variables) to store the persistent data of the contracts, and the state variables in turn are recorded in the blockchain. The \emph{contract state} then can be formally defined as the state variable set $ V = \{V_{1}, V_{2}, V_{3},...,V_{N}\} $ \cite{bose2022sailfish}. A transaction (\emph{i.e.}, function invocation) can change the state(s) of related smart contracts, and the state updates are permanently recorded in the blockchain \cite{soliditylang}. As the blockchain storage space is limited, typically only critical data (\emph{e.g.}, contract owner's address) are stored using state variables.

To further control the visibility and access rights of state variables, Solidity provides various modifiers  \cite{fang2023beyond}. Through these modifiers, developers can precisely control the access scope and modification rules of variables, thereby enhancing the security and transparency of contracts. Visibility modifiers such as \texttt{public}, \texttt{internal}, and \texttt{private} can control the access scope of variables. Variables modified by \texttt{public} can be accessed externally, variables modified by \texttt{internal} can only be accessed within a contract or in an inherited contract, and variables modified by \texttt{private} can only be accessed in the contract that declares it. In addition, \texttt{constant} and \texttt{immutable} modifiers control the mutability of variables. The state variables modified by both \texttt{constant} and \texttt{immutable}
cannot be modified after the contract has been constructed. 
Also, these two modifiers can help save gas because the corresponding variable values are compiled directly into the bytecode without occupying the EVM storage space \cite{soliditylang}. 

\subsection{Inconsistent State Update Bugs}
\label{bugdefinition}
Smart contracts typically have numerous state variables, and these variables are often correlated such that modifications to one variable typically necessitate synchronized updates of other related variables \cite{torres2021confuzzius}. However, when the developers conduct updates to one variable, they may forget to update the correlated variables or wrongly update the correlated variables. 

\textbf{Definition (Inconsistent state update bug).} \emph{Given a smart contract $C$ that ideally has a set $V = \{V_{1},...,V_{m}\} $ of correlated state variables that necessitate synchronized updates, if a subset $S$ ($S \subsetneq V \land S\neq \emptyset $) of state variables have been updated, but the set $ V\setminus S$ of state variables have been incorrectly updated, then the contract $C$ is said to have an inconsistent state update bug. }

Note that inconsistent state update bugs differ from state-inconsistency bugs \cite{bose2022sailfish,fse25}. The latter mainly arises from uncertain transaction orders or hijacked control-flow inside a transaction \cite{qi2023smart, su2020solution}, ultimately leading to state discrepancies. Instead, the inconsistent state update bug emphasizes inherent defects in the contract's internal state update logic rather than state deviations caused by external environmental factors.


\emph{Example.} Fig.~\ref{fig:Background} gives a real‑world example of an inconsistent state update bug. The corresponding contract is used to convert user payments into Non-Fungible Tokens (NFTs) and update the credit balance.
In the \texttt{\_processPayment()} function that handles user payments, the \texttt{creditsOf} mapping serves as the core state variable for maintaining users' credit balances, and the \texttt{\_mintBestAvailableTier()} function is called to convert the payment into a NFT. Users have the option to decide whether to convert their payment into a NFT immediately based on their individual needs. 
If the user wishes to mint the NFT immediately, the \texttt{\_dontMint} parameter can be set to false, allowing the function to proceed with the subsequent execution of the \texttt{\_mintBestAvailableTier()} function to mint the NFT and then update the credit balance.
If a user wishes to make a payment now and defer minting the NFT, they may set the \texttt{\_dontMint} flag to true, which causes the function to return early within the \texttt{if(\_dontMint)} conditional block.
This implementation fails to modify the \texttt{creditsOf} mapping, 
resulting in an inconsistent state update. The absence of this partial status update directly leads to the loss of the user's payment amount. An effective solution to this issue is to insert \texttt{creditsOf[\_data.beneficiary] += \_value;} prior to the return statement. This approach ensures accurate credit balance accounting regardless of minting selections, thereby maintaining state update consistency.

Inconsistent state update bugs are common in practice, and a recent study shows that they account for 18\% of the total Code4rena bugs \cite{zhang2023demystifying}. Thus, proper identification of correlations between state variables and achieving synchronized updates constitute a fundamental task in smart contract development. 
Depending on the specific state variables involved, the consequences of inconsistent state update bugs rang from wrong statistics to financial losses. 
In fact, several well-known smart contract attacks (such as the DAO attack, the Parity wallet hack, \emph{etc.}) are closely related to inconsistent state update bugs \cite{brent2020ethainter, perez2021smart}. Specifically, the DAO attack falls under the category of Interim State Exploit in the classification of exploitation methods presented in Sec.~\ref{secExploitation} of this paper. Worse still, vulnerability attacks using inconsistent state update bugs are on the rise and account for 11\% of the real-world exploits \cite{zhang2023demystifying}. For instance, three recent exploits have resulted in around \$3.8 million loss \cite{exploit1,exploit2,exploit3}. Given this, it is of vital importance for avoiding inconsistent state update bugs that are exploitable and cause asset losses, and we hereafter refer to these bugs as inconsistent state update vulnerabilities for simplicity. This work focuses exclusively on inconsistent state update vulnerabilities to enable a thorough and comprehensive understanding of them.

\begin{figure}[htbp]
\centering
\includegraphics[width=0.7\textwidth]{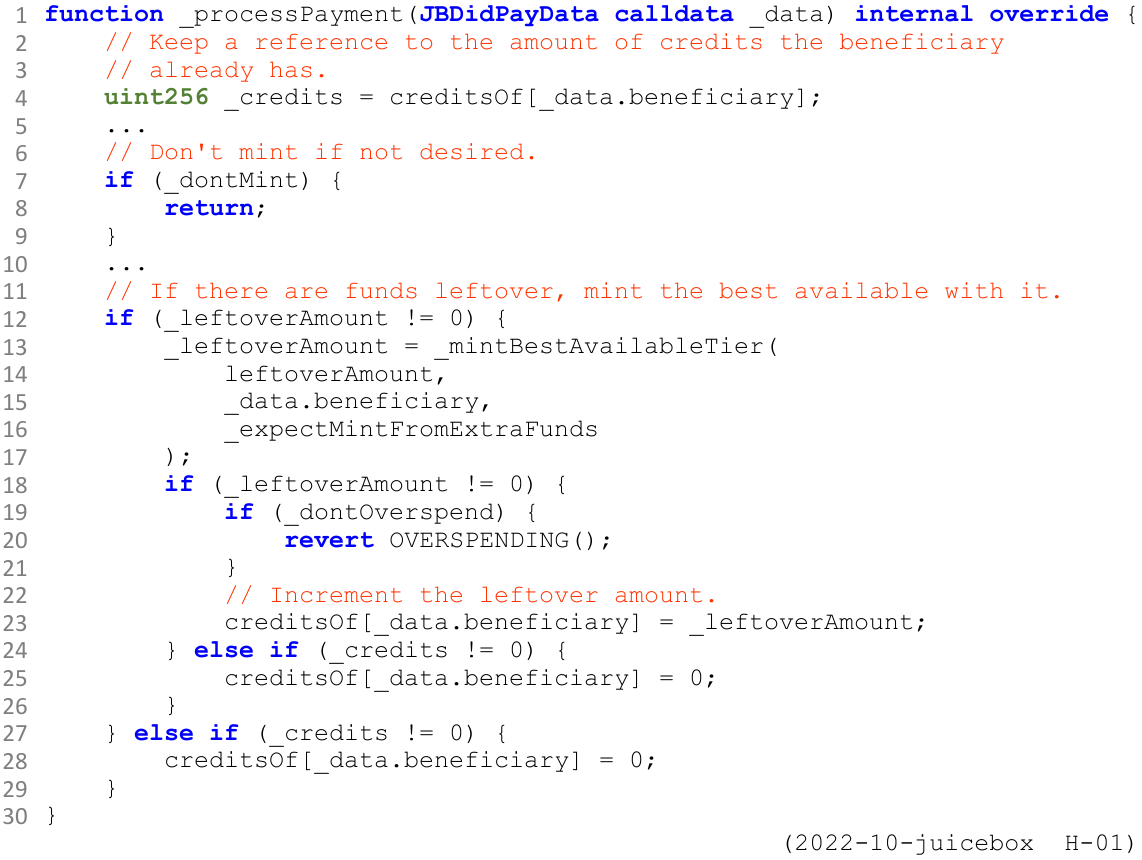}
\caption{A real‑world example of an inconsistent state update bug (\url{https://code4rena.com/reports/2022-10-juicebox}).}
\Description{A real‑world example of an inconsistent state update bug.}
\label{fig:Background}
\end{figure}

\section{Methodology}
\label{Methodology}
This section describes our methodology to perform our original study of inconsistent state update vulnerabilities in smart contract.
\subsection{Subjects and Data Used in Our study}
\label{SecData}
We collect the dataset of inconsistent state update vulnerabilities from Code4rena \cite{Code4rena}, the largest and most reputable decentralized audit competition platform \cite{sun2024gptscan, xi2024pomabuster, wang2024contracttinker, liu2024using}. Each competition held by the Code4rena platform lasts 3 to 7 days, and aims to audit real-world DeFi (decentralized finance) projects to ensure the security of the projects before they are officially deployed. Individuals, companies, and institutions from all over the world can participate in the competition. After the competition, Code4rena's jury (composed of experienced audit experts selected by the community) reviews the bug reports with the project development team, confirming the real bugs. In addition, a criticality level will be assigned to each bug, which can be low, medium, or high. High-level bugs can cause asset losses, and thus be exploited by attackers (\emph{i.e.}, they are vulnerabilities). Depending on the level of criticality of the reported bug and also the number of simultaneous reports for the bug, the development team of each theme project will provide a bonus ranging from 20,000 to 1 million US dollars as a reward. To date, the platform has paid more than 20 million dollars in rewards, demonstrating the representativeness and reliability of the bugs of the platform. Note that the bugs found on the Code4rena platform are typically difficult for machines to automatically audit \cite{zhang2023demystifying}. In other words, they are beyond the capability of existing tools.

The Code4rena platform provides the source code snippets, bug reports, and relatively authoritative identification of real-world smart contract bugs. Currently, there is a lack of comparable auditing platforms of similar scale and credibility, making Code4rena the most suitable—and indeed the only—data source for this study. Furthermore, as the Code4rena platform assigns criticality level to bugs, which facilitates the identification of targeted inconsistent state update vulnerabilities in this work. More specifically, the assigned  high-level bugs are vulnerabilities 
that are exploitable and cause asset losses. 
Even though inconsistent state updates might exist in lower-risk vulnerabilities, they do not lead to economic loss. Thus, we only selected the categories authoritatively marked as high-risk by the Code4rena platform, which aligns with the approach taken by Zhang and Wang \emph{et al.} \cite{zhang2023demystifying, wang2024contracttinker}.


We now give the data collection process in detail. We first collected audit reports from a total of 352 contests (each contest targets at a real-world smart contract project) hosted on the Code4rena platform between April 2021 and October 2024. 
For each report, we extracted all bugs in the report that were annotated with \emph{High Risk Finding} (\emph{i.e.}, the assigned high-level bugs), resulting in 1,361 vulnerabilities.
We then conducted a detailed manual review of the descriptions of these 1,361 vulnerabilities provided in the reports. Each vulnerability was independently evaluated by two authors of this paper according to the formal definition given in Sect.~\ref{bugdefinition}. 
Any disagreement will be turned to the third author. We reach consensus for all vulnerabilities after the new researcher gets involved.
To reduce the subjectiveness threat, three authors conducting this evaluation possess over 3 years of development and research experience in Solidity smart contracts, and they thoroughly analyzed the source code, bug reports, code comments, and fix suggestions of the 1,361 vulnerabilities.
As a final result of this process, we identified 116 vulnerabilities that fall into this category, which serve as the subjects of analysis in this study. We calculated the complexity metrics for the contract projects to which these 116 vulnerabilities belong. 
The average LLOC (logical lines of code, without comments and white spaces) \cite{ajienka2020empirical} is 14603.51, the average NF (number of functions) is 1703.14, and the average number of state variables is 683.67.

\subsection{Study Methodology}
\label{StudyMethod}
For each inconsistent state update vulnerability, we thoroughly analyze the full source code (which is available through GitHub platform), code comments, bug reports, and fix suggestions of the relevant contracts to fully understand the root causes for the vulnerabilities, the effective fix strategies for the vulnerabilities, and the exploitation methods for the vulnerabilities. To minimize our own subjective judgment, each vulnerability is analyzed independently by two authors of the paper. Subsequently, we classify and describe the vulnerabilities in three dimensions: the root cause, the fix strategy, and the exploitation method. It is worth mentioning that our researchers have at least 3 years of experience in Solidity smart contract development and research. 

The manual check is in particular composed of two phases. In the first phase, two independent authors of the paper derive an initial result after the analysis. 
After the classification, the Cohen’s kappa coefficient is calculated \cite{cohen1960coefficient}, which is a statistic for measuring inter-rater reliability for qualitative items. The Cohen’s kappa coefficients are 0.82, 0.84, and 0.88 for the results about the root cause, the fix strategy, and the exploitation method respectively, suggesting that the agreement rate is good.
In the second phase, the two authors compare the initial results derived, have a discussion about the disagreements, and refine their initial results according to the outcome of the discussion (\emph{i.e.}, the consensus that has been reached). Note that Cohen’s kappa coefficients and multi-phase manual checks have been widely used by existing works \cite{li2023understanding, ICSE43902202100019, 1011453434279, yuannotation}.



\section{Root Causes for Inconsistent State Update Vulnerabilities (RQ1)}
\label{causes}
This section aims to answer RQ1: ``What are the root causes of inconsistent state update vulnerabilities?'' To systematically identify the root causes and categorize the vulnerabilities, we conducted a meticulous qualitative analysis of the 116 vulnerability samples described in Sec.~\ref{SecData}. The specific approach involved performing an in-depth review of the source code, code comments, bug reports, and fix suggestions for each vulnerability to  identify the core defects leading to the state inconsistency. Based on the analysis of defect patterns, we constructed a taxonomy of root causes comprising four categories, as shown in Table~\ref{reason}. To ensure classification consistency, we followed the multi-person independent labeling and discussion process detailed in Sec.~\ref{StudyMethod}. The following subsections elaborate on the root causes.


\begin{table}
\caption{Category of the root causes for inconsistent state update vulnerabilities.}
\begin{center}
\begin{tabular}{ l c c }
\toprule
\textbf{Root cause categories} & \textbf{Instance} & \textbf{Percentage}  \\
\midrule
\makecell[l]{Dynamic Dependent Update Omission} & 55 & 47.41\% \\
Incorrect Logic Update & 40 & 34.48\% \\
Variable Omission & 12 & 10.34\% \\
Initialization/Re-initialization Omission & 9 & 7.76\% \\
\bottomrule
\end{tabular}
\label{reason}
\end{center}
\end{table}

\subsection {Dynamic Dependent Update Omission (Omission of Update for Dynamic Operation Dependent State Variables, 47.41\%).}
This category of inconsistent state update vulnerabilities occurs for transactions that contain multiple operation steps. Each operation of a smart contract often affects multiple state variables, and the real-time nature of state variables is the basis for the safe operation of smart contracts. However, developers often fail to develop a comprehensive state variable update strategy and ignore state management, resulting in the failure of smart contracts to update dependent state variables that reflect dynamic states in a timely manner after executing operations \cite{codebyankita, KaiaDocs}. Once these state variables are not updated correctly, the contract state will not match the actual operation, causing exploit opportunities.

Fig.~\ref{fig:Price} illustrates a real-life inconsistent state update vulnerability,
and the corresponding contract is designed to track and synchronize the market price of the VADER token through a Time-Weighted Average Price (TWAP) mechanism. The \texttt{setupVader()} function is executed during the contract deployment to set the initial price of VADER and store it in the \texttt{previousPrices} array, which is intended to track the historical price values of the token in subsequent transactions. Then, by calling the \texttt{\_addVaderPair()} function, the VADER token is paired with its related assets and an external oracle, completing the token’s initialization configuration. Following this, the \texttt{\_updateVaderPrice()} function is executed. In this function, 
the \texttt{nativeTokenPriceAverage} is calculated to synchronize the VADER token's TWAP with real-time market conditions. 
{Additionally, the \texttt{currentLiquidityEvaluation} is calculated based on the current asset reserves in the liquidity pool and the value of \texttt{previousPrices}, and its calculation is intended to assess liquidity.}
While the cumulative price data is gathered and updated, the \texttt{previousPrices} remains fixed at the initial VADER price and is not updated during each synchronization, whereas all other variables reflect the latest market conditions. Therefore, the liquidity evaluations always use the initial price, causing a mismatch between the contract’s asset valuation and the actual market conditions, resulting in price distortion. 

\begin{figure}[htbp]
\centering
\includegraphics[width=0.65\textwidth]{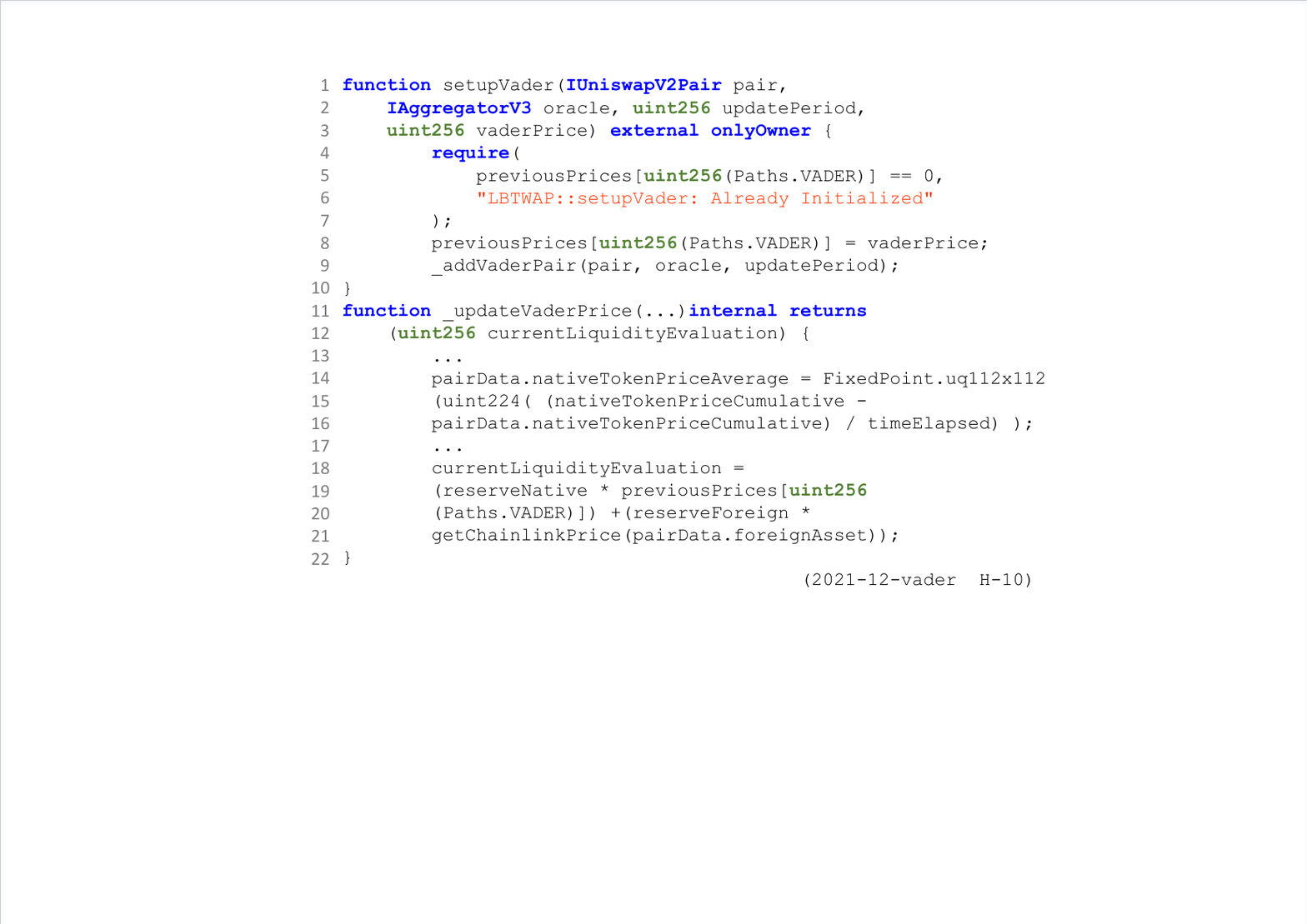}
\caption{A vulnerability in a liquidity-weighted price oracle contract (\url{https://code4rena.com/reports/2021-12-vader}).}
\Description{A price manipulation vulnerability.}
\label{fig:Price}
\end{figure}

For vulnerabilities with this cause category, they happen frequently for \emph{state synchronization in permission management}, with 12.07\% of all the 116 vulnerabilities occurring for this scenario. This type of inconsistent state update vulnerability exists in the authorization mechanism and stems from the fact that the state variables related to permission control are not updated in time after key operations, resulting in a disconnect between permission management and the actual logic.
In complex authorization operations, the old authorization state is not cleaned up when it comes to historical state cleanup.
Fig.~\ref{fig:Permission} illustrates an inconsistent state update vulnerability in a digital asset ownership contract.
The \texttt{approve()} function grants a specific address the authority to manage the token, and it particularly sets \texttt{getApproved[\_tokenId]} to the authorized address. Once authorized, the address is permitted to perform token transfer operations within the \texttt{\_transfer()} function. Subsequently, when the \texttt{\_transfer()} function is called, it first verifies the caller’s authorization through \texttt{\_requireAuthorized()}, and then executes the actual transfer operation via \texttt{transferPositionEEC7A3CD()}.
However, since \texttt{\_transfer()} does not revoke the authorization record in \texttt{getApproved[\_tokenId]} after the transfer, the originally approved address retains its permissions. This allows this address to repeatedly pass \texttt{\_requireAuthorised()} checks and call \texttt{\_transfer()} to reclaim the tokens from the new owner, resulting in unauthorized token recovery.

\begin{figure}[htbp]
\centering
\includegraphics[width=0.65\textwidth]{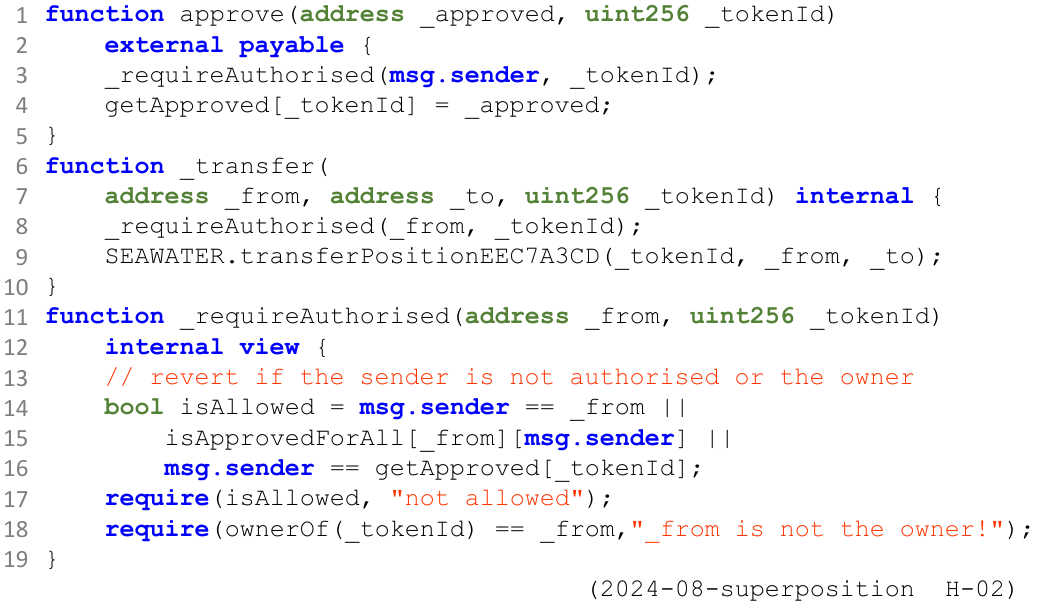}
\caption{A vulnerability in a digital asset ownership contract (\url{https://code4rena.com/reports/2024-08-superposition}).}
\Description{A state synchronization bug in permission management.}
\label{fig:Permission}
\end{figure}

\vspace{1mm}
\begin{center}
\fcolorbox{black}{gray!25}{\parbox{0.97\linewidth}{
\textit{\noindent\textbf{Finding 1}: 
Around half of inconsistent state update vulnerabilities arise due to ``Dynamic Dependent Update Omission'', which means for transactions involving multiple operation steps, some state variables fail to be updated in time after each operation step. In particular, this category of vulnerabilities occurs frequently for state synchronization in permission management.
This reveals a critical blind spot in existing automated tools: while they excel at analyzing implemented code, they struggle to identify update operations that should exist but were omitted by developers.
}

\textit{
\noindent\textbf{Implication 1}: 
To avoid inconsistent state update vulnerabilities during the smart contract development, developers must ensure atomic updates for state variables involved in transactions that contain multiple operation steps. In particular, 
Solidity-specific features can be employed to establish state synchronization mechanisms. One potential solution is through 
event notifications, which record state changes and timely trigger corresponding compensation or repair mechanisms when state abnormalities are detected.
}}}
\end{center}
\vspace{1mm}

\subsection{Incorrect Logic Update (Conducted State Variable Update with Incorrect Logic, 34.48\%).}
This category of inconsistent state update vulnerabilities arises when performing state variable updates, the underlying logic for the update is incorrect. Due to the incorrect logic, the contract state deviates from the expected state. 
For instance, some state variables (such as counters or allocation ratios) are incorrectly reset or overwritten when the state transitions, resulting in the loss of historical data or state inconsistencies. Additionally, state variables often lack isolation mechanisms in multi-user interactions, making them susceptible to logical errors caused by conflicting operations. Furthermore, when a variable is defined as a memory type but should be a storage type, updates may fail to persist, causing further issues. 
These vulnerabilities can lead to direct financial losses (such as token over-issuance or locked funds), failure of governance systems (such as incorrect voting outcomes), or dysfunction of contract features (such as improper cross-chain message transmission). 

Fig.~\ref{fig:Incorrect-Logic} illustrates an inconsistent state update vulnerability in an oracle-driven asset-backed loan contract.
This contract allows borrowers to use NFTs as collateral to secure loans, with lenders determining whether to issue loans and charge interest based on asset valuations provided by an oracle. The \texttt{updateLoanParams()} function enables lenders to update the loan parameters when the loan status is \texttt{LOAN\_OUTSTANDING}. In this function, a parameter \texttt{params}—a struct containing fields such as \texttt{oracle} and \texttt{duration}—is passed in. All fields in the parameter, except for the \texttt{oracle}, are validated through require statements to prevent lenders from increasing the borrower's risk by adjusting the parameters after the loan has commenced. However, the \texttt{oracle} field is not checked for consistency with the currently registered oracle. This allows a malicious lender to replace the oracle with one they control, while all other parameters are correctly updated (line 18). Subsequently, the \texttt{removeCollateral()} function is responsible for capturing the borrower's collateral when the loan is in default or has matured, according to the value of the \texttt{oracle}. When \texttt{removeCollateral()} is executed, the contract evaluates \texttt{loanParams.oracle.get()}—now potentially pointing to a manipulated oracle—to obtain a falsified asset valuation. This allows the bypassing of collateral value checks (line 27) and facilitates the improper appropriation of the borrower's assets.

\begin{figure}[htbp]
\centering
\includegraphics[width=0.65\textwidth]{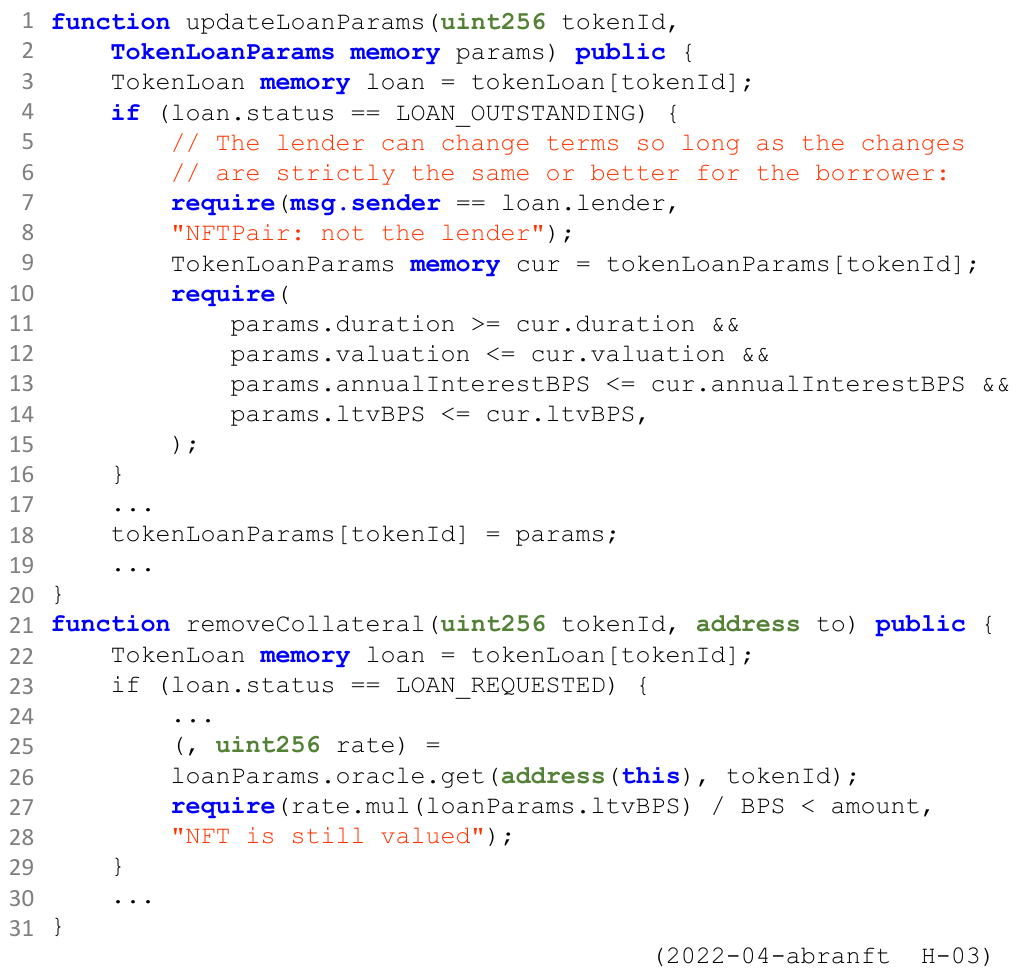}
\caption{A vulnerability in an oracle-driven asset-backed loan contract (\url{https://code4rena.com/reports/2022-04-abranft}).}
\Description{Another price manipulation vulnerability.}
\label{fig:Incorrect-Logic}
\end{figure}

\vspace{1mm}
\begin{center}
\fcolorbox{black}{gray!25}{\parbox{0.97\linewidth}{
\textit{\noindent\textbf{Finding 2}: 
Around one third of inconsistent state update vulnerabilities arise due to ``Incorrect Logic Update'', which means when performing state variable updates, the underlying logic for the update is incorrect. 
This suggests that even if developers remember to update all variables and write values correctly, the complexity of the update logic itself remains a primary threat source. 
}

\textit{
\noindent\textbf{Implication 2}: 
For instructions that conduct state variable update, formally analyzing their data and control flow will be extremely beneficial for detecting inconsistent state update vulnerabilities, e.g., through model checking or symbolic execution. 
}}}
\end{center}
\vspace{1mm}

For vulnerabilities with this cause category, there exist two subtypes that are especially prevalent and we next discuss these two subtypes in more detail.

First, the subtype \emph{Improper Call Sequence}
accounts for 12.07\% of all the 116 studied inconsistent state update vulnerabilities. This subtype of inconsistent state update vulnerabilities lies in the improper design of function call order. 
Fig.~\ref{fig:Reentrancy} illustrates an inconsistent state update vulnerability in an interchain token transfer service contract.
The \texttt{\seqsplit{expressReceiveTokenWithData()}} function is responsible for handling cross-chain token reception and transfers. 
Within this function, \texttt{gateway.isCommandExecuted()} is used to check whether the cross-chain transaction corresponding to the \texttt{commandId} has been executed. If the \texttt{commandId} has been marked as processed, it triggers a revert to prevent the same transaction from being executed again. Then, the function call \texttt{safeTransferFrom()} transfers the token from the caller's account to the target address. Next, the function call \texttt{\_expressExecuteWithInterchainTokenToken()} performs the cross-chain operation, transferring the token from the current chain to the target chain. Finally, the function call \texttt{\_setExpressReceiveTokenWithData()} is employed to update certain critical states after the token transfer is completed. At this point, the transaction state has already been changed, but because the call to \texttt{\seqsplit{\_setExpressReceiveTokenWithData()}} occurs last in the sequence (\emph{i.e.}, it has not yet been executed), causing the \texttt{commandId} to not be marked as executed at this moment.

\begin{figure}[htbp]
\centering
\includegraphics[width=0.72\textwidth]{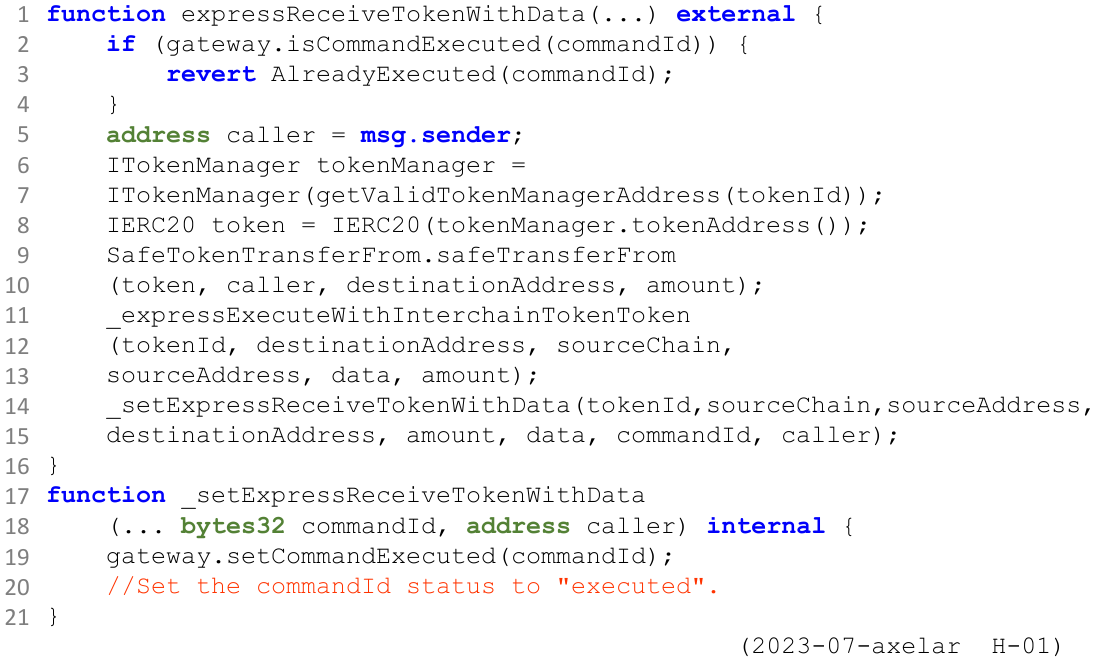}
\caption{A vulnerability in an interchain token transfer service contract (\url{https://code4rena.com/reports/2023-07-axelar}).}
\Description{A reentrancy bug.}
\label{fig:Reentrancy}
\end{figure}

Second, the subtype \emph{Improper Boundary Condition Handling} accounts for 6.03\% of all the 116 studied inconsistent state update vulnerabilities. This subtype of inconsistent state update vulnerability lies in incomplete consideration of the state variable update logic for specific scenarios in contract design. In particular, these vulnerabilities occur in extreme or edge cases (such as an initial balance of zero, cycle transition points, or abnormal inputs) where consistency in state variable updates is not ensured. Although these conditions are rare in regular usage, they can trigger specific logic branches, causing certain state variable updates to be miscalculated. 
From the validation perspective, the commonality of vulnerabilities of this subtype is insufficient condition path coverage, preventing critical state variables from staying in sync with actual operations. Fig.~\ref{fig:Computation} illustrates an inconsistent state update vulnerability caused by improper handling of boundary conditions, and the corresponding contract is used to charge fees based on the time difference when users deposit or withdraw tokens. 
The \texttt{handleFees} function calculates fees based on the time difference (\texttt{timeDiff}), with the variable \texttt{lastFee} recording the timestamp of the last \texttt{handleFees} call. However, when \texttt{startSupply} is 0 (which can happen if all token holders burn their tokens), the \texttt{handleFees} function does not update \texttt{lastFee}. When new users enter again, the \texttt{handleFees} function calculates fees based on the old \texttt{lastFee} while time has progressed. This results in an overcalculation of fees, thereby diluting the shares of existing token holders.

\begin{figure}[htbp]
\centering
\includegraphics[width=0.62\textwidth]{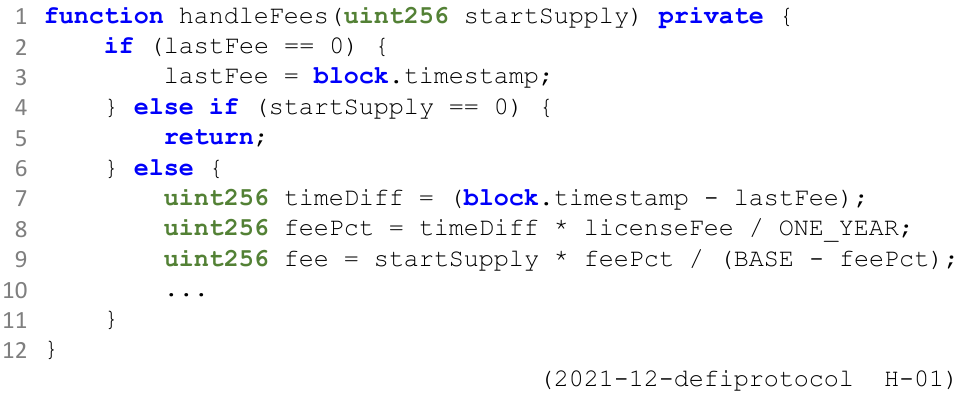}
\caption{A vulnerability in a time-based fee asset vault contract (\url{https://code4rena.com/reports/2021-12-defiprotocol}).}
\Description{A calculation bug.}
\label{fig:Computation}
\end{figure}

\vspace{1mm}
\begin{center}
\fcolorbox{black}{gray!25}{\parbox{0.97\linewidth}{
\textit{\noindent\textbf{Finding 3}: 
For vulnerabilities with the cause category ``Incorrect Logic Update'', two subtypes ``Improper Call Sequence'' and ``Improper Boundary Condition Handling'' are prevalent.
They can be distinctly attributed to two types of programming oversights: one pertains to the high-level execution flow across function calls and the other concerns the low-level conditional branches within functions. This distinction helps to more precisely pinpoint and categorize the challenges developers face at different levels of abstraction.
}

\textit{\noindent\textbf{Implication 3}: 
Guided fuzzing towards call sequence diversity and edge conditions will also be helpful for detecting inconsistent state update vulnerabilities. 
}}}
\end{center}
\vspace{1mm}

\subsection{Variable Omission (Omission of Critical State Variables, 10.34\%).}
This particular category of inconsistent state update vulnerabilities arises from the absence of certain critical state variables in the contract, such as flags, mappings, and counters. Note that ideally the contract has the critical state variable(s), so this category of vulnerabilities aligns with our bug definition in Sect.~\ref{bugdefinition}. 
The impacts of these vulnerabilities include illegal fund transfers, unfair reward distribution, and unpaid debts, directly threatening the economic model of the contract and the interests of users. 

Fig.~\ref{fig:Front-running} illustrates an inconsistent state update vulnerability caused by the absence of a critical distinguishing variable, and
the corresponding contract is designed to manage installment loans and various loan terms. Specifically, the \texttt{refinanceFull()} function allows for the refinancing of an existing loan by updating its terms (such as the principal amount, interest rate, \emph{etc.}).  The \texttt{addNewTranche()}  function enables the borrower to add new installments to the existing loan, thereby expanding the structure of the loan.
Both \texttt{refinanceFull()} and \texttt{addNewTranche()} functions use the same signature verification mechanism \texttt{\_checkSignature}, but lack a key state variable to differentiate between loan restructuring and adding a new loan tranche. 


\begin{figure}[htbp]
\centering
\includegraphics[width=0.76\textwidth]{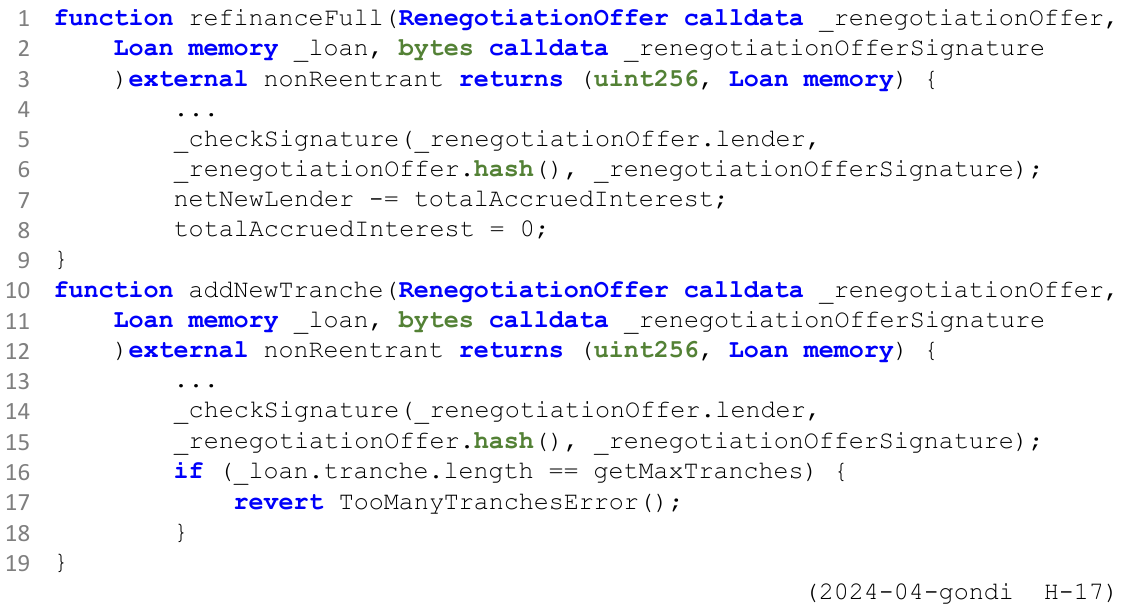}
\caption{A vulnerability in a flexible loan contract (\url{https://code4rena.com/reports/2024-04-gondi}).}
\Description{A front-running bug.}
\label{fig:Front-running}
\end{figure}

\subsection{Initialization/Re-initialization Omission (Omission of Initializations/Re-initializations of Critical State Variables, 7.76\%).}
This particular category of inconsistent state update vulnerabilities arises due to the absence of explicit initializations/re-initializations of certain critical state variables in the contract, leading to incomplete or erroneous state logic. Again, note that other correlated state variables have been correctly updated, thus this category of vulnerabilities aligns with our bug definition in Sect.~\ref{bugdefinition}. Due to the absence of explicit initializations/re-initializations, default/old values will instead be used, which in turn can lead to  misjudgments in certain scenarios. For example, the default value of a Solidity timestamp (a \texttt{uint} type) is 0, which is often mistakenly interpreted as no event trigger condition being met. Additionally, dynamically structured data (such as arrays or mappings) are initialized as empty by default, which may prevent functions from handling edge cases properly. 

Fig.~\ref{fig:Initialization} illustrates an inconsistent state update vulnerability caused by omission of re-initializations of critical state variables, and the corresponding contract is used for proposal validation and dispute resolution. Specifically, the \texttt{initLPP()} function is used to set the metadata of a proposal (including data such as \texttt{\_uuid}, \texttt{\_partOffset}, \emph{etc.}). The function directly reads the existing \texttt{metaData} of a proposal from storage and updates \texttt{\_partOffset} and \texttt{\_claimedSize} based on it before writing it back. This approach reuses the existing state rather than re-initializing a fresh struct, causing exploit opportunities.


\begin{figure}[htbp]
\centering
\includegraphics[width=0.7\textwidth]{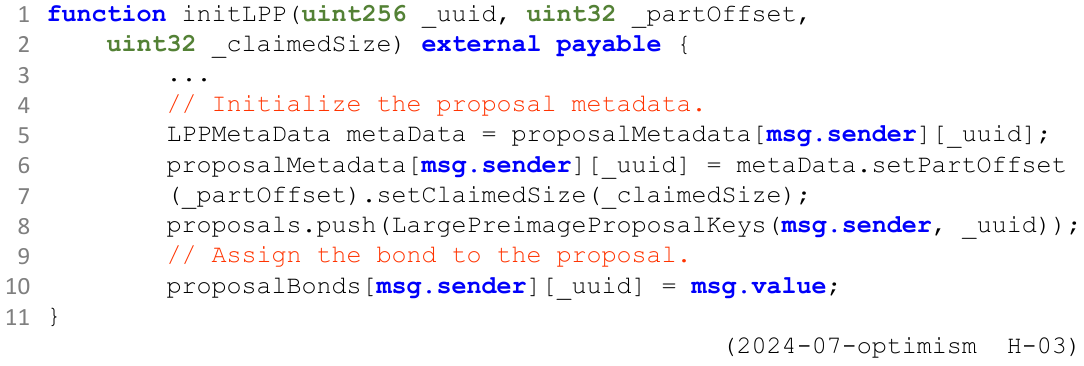}
\caption{A vulnerability in a proposal validation and dispute resolution contract (\url{https://code4rena.com/reports/2024-07-optimism}).}
\Description{A repeated transaction bug.}
\label{fig:Initialization}
\end{figure}

\vspace{1mm}
\begin{center}
\fcolorbox{black}{gray!25}{\parbox{0.97\linewidth}{
\textit{\noindent\textbf{Finding 4}: 
Around one fifth of inconsistent state update vulnerabilities belong to two particular cause categories ``Variable Omission'' and ``Initialization/Re-initialization Omission'', which mean the absence of certain critical state variables and the absence of explicit initializations/re-initializations of certain critical state variables in the contract respectively.
Overall, developers tend to overlook state management at specific stages of the smart contract lifecycle (such as deployment and reset). This arises particularly due to habitual reliance on Solidity's default initial values, but these values often fail to meet business logic expectations.
}

\textit{
\noindent\textbf{Implication 4}: 
To detect the absence of critical state variables, a potential method is using cross-checking which collects functionally similar program slices and infers missed state variables
through majority-voting. 
Besides, IDEs could highlight uninitialized state
variables for further check, ideally on top of advanced call analysis techniques.
}}}
\end{center}
\vspace{1mm}

\section{Fix Strategies for Inconsistent State Update Vulnerabilities (RQ2)}
\label{fix}
This section is dedicated to answering RQ2: ``How can developers fix inconsistent state update vulnerabilities to avoid causing harm?'' Since these vulnerabilities were identified during the pre-deployment audit phase, our analysis is based on the discussed and validated fix recommendations available on the Code4rena platform. We conducted an in-depth manual analysis of the audit reports, community discussions, and suggested fixes related to each vulnerability to extract and summarize the fix strategies that developers should adopt. To gain a deeper understanding of the intrinsic relationship between fix strategies and root causes, we further calculated the \emph{lift} value, 
which quantifies the strength of their association. The subsequent content of this section first introduces the fix strategies for inconsistent state update vulnerabilities and their distribution, followed by a discussion of the correlation between the fix strategies and the root causes.

\subsection{Categories of Fix Strategies}

According to the vulnerability fix strategies, the 116 inconsistent state update vulnerabilities can be classified into four categories, as shown in Table~\ref{fixTable}. Since some vulnerabilities require multiple fix strategies, they may belong to multiple categories simultaneously. 

\begin{table}
\caption{Category of the fix strategies for inconsistent state update vulnerabilities.}
\begin{center}
\begin{tabular}{ l c c }
\toprule
\textbf{Fix strategy categories} & \textbf{Instance} & \textbf{Percentage}  \\
\midrule
Direct Variable Change & 68 & 58.62\% \\
Redesign Algorithm/Data Structure & 27 & 23.28\% \\
Reorder Sequence & 17 & 14.66\% \\
Change Conditions & 12 & 10.34\% \\
\bottomrule
\end{tabular}
\label{fixTable}
\end{center}
\end{table}

\subsubsection{Direct Variable Change (Directly Modify Computations on the Unsafe State Variables, 58.62\%).} 
This fix involves directly adding, removing, and changing operations on unsafe state variables, aiming to ensure that the contract is not affected by illegal inputs or erroneous calculations during execution. Developers believe that state variables may be influenced by external inputs, overflow risks, or malicious manipulation, leading to errors or security vulnerabilities. Thus, fix strategies often involve timely updates of critical data variables, adding overflow protection, removing unsafe operations, or performing secure type conversions to improve computational reliability. 
As the search space for these fix strategies is relatively small, there exists the prospect of well-designed tools to automatically conduct these changes. Given much progress has been made in recent years towards automatically detecting \cite{multiple-fault,guifault,yufse, flsurvey, yuguifl} and fixing \cite{6035728, monperrus2018automatic, yuEmSE, yutse,yang2025parameter,urli2018design,9393494,yangEMSE,xue2025} relatively simple bugs, this prospect is huge.
To fix the vulnerability in Fig.~\ref{fig:Price}, after synchronizing the price (\emph{i.e.}, calculating \texttt{nativeTokenPriceAverage}), \texttt{previousPrices[uint256(Paths.VADER)]} should be immediately updated to the latest value to ensure the timeliness of liquidity. Furthermore, to fix the vulnerability in Fig.~\ref{fig:Permission}, a statement \texttt{getApproved[\_tokenId] = address(0);} should be added after the transaction in the \texttt{\_transfer()} function to prevent the old approved address from abusing its permissions.
Finally, to fix the vulnerability in Fig.~\ref{fig:Initialization}, multiple remediation strategies need to be implemented simultaneously, and one key strategy is to remove the unsafe operation that assigns values to \texttt{metaData} using obsolete storage (line 5).

For vulnerabilities with this fix strategy, the targeted state variables are frequently time-related  (12.93\% of all the 116 vulnerabilities occurred for this scenario),
as timestamps and deadlines often act as critical triggers for state transitions and financial computations. A single outdated value can disrupt core mechanisms like fee accumulation, reward distribution, and access control periods.
The large percentage of time-related variables necessitates a strict definition and management of each lifecycle phase during the contract’s design and implementation to ensure that every transition is secure and behaves as expected.
To fix the vulnerability in Fig.~\ref{fig:Computation}, \texttt{lastFee} should be updated to the current time \texttt{block.timestamp} when \texttt{startSupply} is zero. This ensures that the protocol can correctly resume fee accrual during the transition from a ``token reset'' state to normal operation.

\vspace{1mm}
\begin{center}
\fcolorbox{black}{gray!25}{\parbox{0.97\linewidth}{
\textit{\noindent\textbf{Finding 5}: 
Around sixty percentage of the studied inconsistent state update vulnerabilities are fixed by directly modifying computations on the unsafe state variables, in particular time-related state variables.
}

\textit{
\noindent\textbf{Implication 5}: The high percentage of this relatively simple fix strategy suggests that it is promising to develop fully automated or semi-automated techniques to fix inconsistent state update vulnerabilities in Solidity smart contracts.
}}}
\end{center}
\vspace{1mm}

\subsubsection{Redesign Algorithm/Data Structure (23.28\%). }
This fix strategy usually involves changes in algorithms or data structures to improve the security and stability of the contract. Developers believe that existing algorithms or data structures have defects. When addressing such vulnerabilities, developers usually need to re-examine the core logic of the contract and adjust the data structure or algorithm to ensure that data storage, retrieval, and updates can be handled reasonably. 
By selecting appropriate \emph{mappings}, \emph{arrays}, \emph{linked lists}, \emph{structs}, \emph{hash tables}, and other data structures to improve the logic of the original algorithm (\emph{e.g.}, setting status flags or designing dynamic calculation mechanisms) ensure the synchronization of states between different operations, and prevent attackers from taking advantage of inconsistent states for manipulation (such as repeated withdrawals, incorrect calculation of funds, \emph{etc.}). These improvements at the algorithm and data structure levels help ensure the consistency of contract states in specific scenarios. 
To fix the vulnerability in Fig.~\ref{fig:Front-running}, the {\texttt{RenegotiationOffer} struct} should include an explicit identifier to distinguish different operation types, ensuring that functions \texttt{refinanceFull()} and \texttt{addNewTranche()} cannot share the same signature.

\subsubsection{Reorder Sequence (Reorder Function Call Sequence or State Variable Update Sequence, 14.66\%).}
This fix strategy lies in adjusting the call order of functions or the sequence of state variable updates to ensure that the contract uses the correct state information during execution. 
For example, when transferring funds or changing permissions, the fix requires updating balance or permission states before executing these operations to ensure that subsequent actions are not affected by outdated state information. Additionally, in cases where multiple state updates are required, the fix often adjusts the placement of these updates to prevent state inconsistencies or information leakage, ensuring that each operation is executed at the right moment. Through these adjustments, the correctness and security of the contract logic can be maintained, preventing vulnerabilities caused by improper execution order. 
In practice, this reordering is often guided by the Checks-Effects-Interactions (CEI) pattern, which provides a robust defense mechanism (particularly against reentrancy attacks).
More specifically, to fix the vulnerability in Fig.~\ref{fig:Reentrancy}, the function \texttt{\_setExpressReceiveTokenWithData()} should be called before any external calls, ensuring that \texttt{commandId} is marked as executed before proceeding with the external transfer.

\subsubsection{Change Conditions (10.34\%).}
This fix strategy typically involves adjusting the execution conditions to prevent or mitigate the impact of inconsistent state update vulnerabilities.
For example, in the implementation of contract functions such as state transitions and permission verification, failure to properly handle edge cases or insufficient condition validation may lead to inconsistent state update vulnerabilities. To address these problems, it is often necessary to introduce or refine conditional checks in the code to ensure consistency in state transitions and enhance logical rigor. 
Overall, this fix strategy underscores the importance of thorough condition validation before state updates.
More specifically, to fix the vulnerability in Fig.~\ref{fig:Incorrect-Logic}, a validation predicate \texttt{params.oracle == cur.oracle} should be added to the \texttt{require} clause in \texttt{updateLoanParams()} to check the oracle field.

\vspace{1mm}
\begin{center}
\fcolorbox{black}{gray!25}{\parbox{0.97\linewidth}{
\textit{\noindent\textbf{Finding 6}: 
Around forty percentage of the studied inconsistent state update vulnerabilities are fixed by three strategies other than directly modifying computations on the unsafe state variables, which usually require deep understanding of program semantics and programmers need to consider correctness, performance and other issues to decide the most appropriate fix strategy.
}

\textit{
\noindent\textbf{Implication 6}: 
There is no silver bullet for fixing inconsistent state update vulnerabilities, debugging tools should provide more bug manifestation information to help developers diagnosis and fix them.
}}}
\end{center}
\vspace{1mm}

\textbf{Noteworthy combination patterns of fix strategies.} Of all the 116 inconsistent state update vulnerabilities, 7 vulnerabilities (approximately 6\%) require multiple fix strategies. Among these, 1 vulnerability requires three fix strategies simultaneously, and 6 vulnerabilities require two fix strategies. A notable pattern is that the fixes for all the 7 vulnerabilities include the ``change conditions'' strategy. Moreover, 5 of them simultaneously combine the ``direct variable change'' and ``change conditions'' strategies. The emergence of this pattern is not accidental: it reveals the intrinsic characteristics of these complex vulnerabilities. More specifically, they involve an incomplete access logic (fixed by the ``change conditions'' strategy) and an update defect that leads to contamination or calculation errors of core state variables (thus requiring the ``direct variable change'' strategy). This indicates that in smart contracts involving multi-layer verification and complex state transitions, defects in conditional logic and defects in state updates can become tightly coupled. Therefore, a single fix strategy is insufficient to completely resolve the issue and coordinated multiple strategies are necessary.
The existence of this pattern has important implications for the development of automated repair tools: rather than being limited to providing single-type patch suggestions, an ideal repair system should possess the ability to identify the composability of vulnerabilities and generate combined fix solutions.

\subsection{Correlation between the Root Causes and Fix Strategies}
\label{Correlation between Root Causes and Fix Strategies}

\begin{figure}[htbp]
\centering
\includegraphics[width=0.80\textwidth]{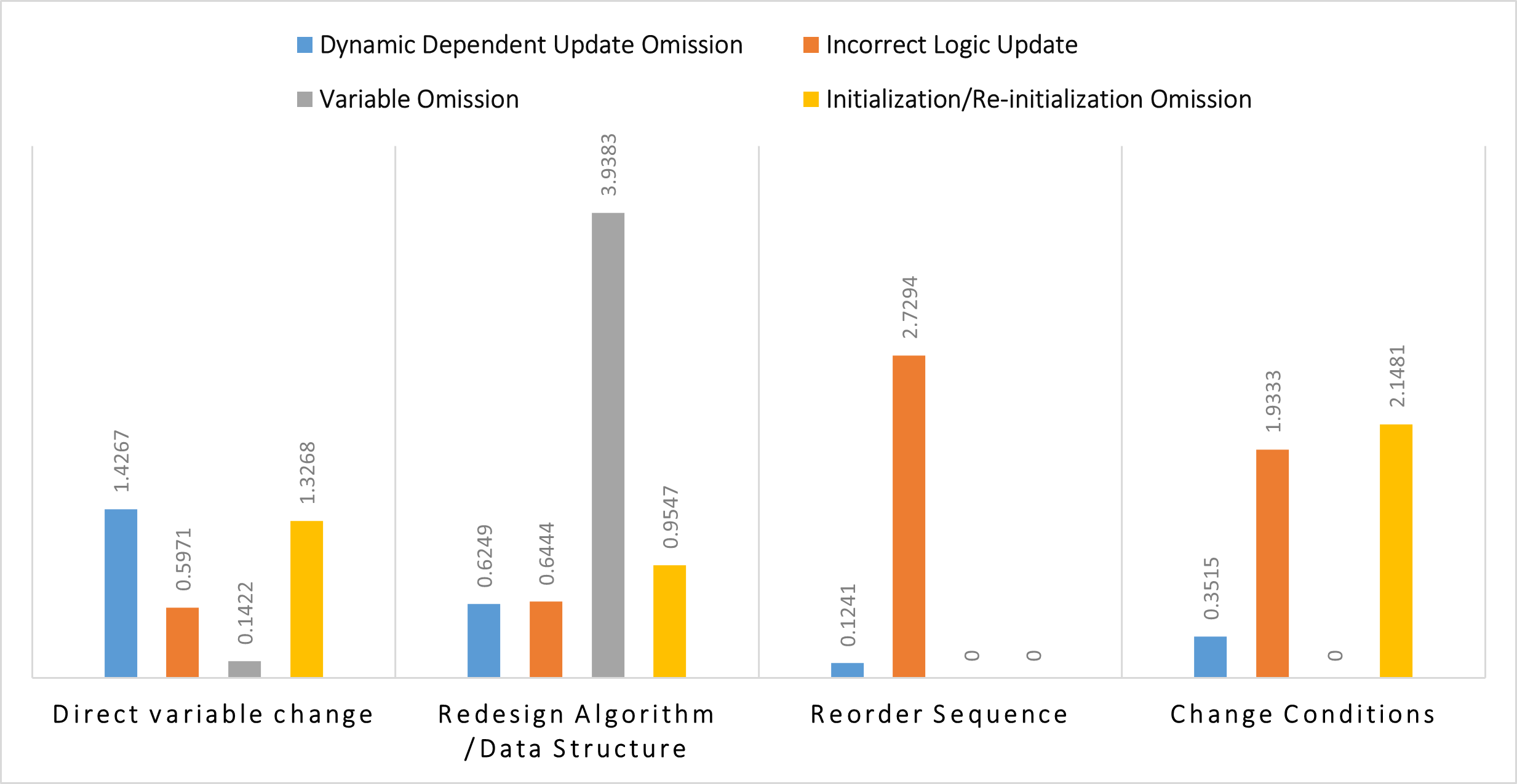}
\caption{The statistical metric \emph{lift} between the root cause of an inconsistent state update vulnerability and its fix strategy.}
\Description{The statistical metric \emph{lift} between the root cause of an inconsistent state update bug and its fix strategy.}
\label{lift}
\end{figure}


To understand the relationship between the cause of an inconsistent update vulnerability and its fix, following previous empirical studies about software bugs (\cite{10.1145/1181309.1181314,10.1145/3297858.3304069,performance}), we calculate a statistical metric named \emph{lift}. For a root cause category \emph{A} and a fix strategy category \emph{B}, the \emph{lift} of them is denoted by \emph{lift(AB)} and calculated as \emph{lift(AB)} = $\frac{P(AB)}{P(A)P(B)} $, where \emph{P(AB)} denotes the probability that an inconsistent state update vulnerability is caused by \emph{A} and fixed by \emph{B}. If the calculated value \emph{lift} is equal to 1, then the root cause \emph{A} is independent with the fix strategy \emph{B}. If the calculated \emph{lift} value is larger than 1, then the root cause \emph{A} and the fix strategy \emph{B} are positively correlated. The positive correlation means that if an inconsistent state update vulnerability is caused by \emph{A}, then it is more likely to be fixed by \emph{B}. If the calculated \emph{lift} value is smaller than 1, then \emph{A} and \emph{B} are negatively correlated.

The result is given in Fig.~\ref{lift}. From this figure, we can see that 1) for the root cause category ``Dynamic Dependent Update Omission'', it has the strongest correlation with fix strategy category ``Direct Variable Change''; 2) for the root cause category ``Incorrect Logic Update'', it has strong correlations with fix strategy categories ``Reorder Sequence'' and ``Change Conditions''; 3) for the root cause category ``Variable Omission'', it has the strongest correlation with fix strategy category ``Redesign Algorithm/Data Structure''; 4) for the root cause category ``Initialization/Re-initialization Omission'', it has strong correlations with fix strategy categories ``Direct Variable Change'' and ``Change Conditions''.  Overall, we can see that there exists a high correlation between causes and fixes in inconsistent state update vulnerabilities. The high correlation of vulnerability causes and fix strategies suggests fruitful revenue in investigating automated fixing tools for inconsistent state update vulnerabilities in smart contracts.

\vspace{1mm}
\begin{center}
\fcolorbox{black}{gray!25}{\parbox{0.97\linewidth}{
\textit{\noindent\textbf{Finding 7}:
The fix strategies of inconsistent state update vulnerabilities in our study are highly correlated with the vulnerability causes.
}

\textit{
\noindent\textbf{Implication 7}: The specific correlation between each vulnerability cause and the fix strategy gives programmers great hints when fixing inconsistent state update vulnerabilities, and this correlation can also guide the design of automated fixing tools. 
}}}
\end{center}
\vspace{1mm}

\section{Exploitation Methods for Inconsistent State Update Vulnerabilities (RQ3)}
\label{exploit}
This section aims to answer RQ3: ``In what forms are inconsistent state update vulnerabilities exploited?'' 
The bug reports from Code4rena pinpoint the vulnerabilities, with many reports providing preliminary descriptions of potential exploitation methods. Building on the bug report, we further inferred and constructed the detailed exploitation methods for the 116 vulnerabilities through an in-depth analysis of source code and code comments. 
These exploitation methods reveal the highly probable attack paths that would be adopted by attackers if the vulnerabilities were not discovered and fixed in time. Furthermore, by analyzing the \emph{lift} values between the root cause categories and the exploitation method categories, we examined the strength of association between root causes and their exploitation methods. The following subsections will elaborate on the distribution proportions, operational mechanisms, and typical scenarios of these exploitation methods.

\subsection{Categories of Exploitation Methods}
\label{secExploitation}

According to the vulnerability exploitation methods, the 116 inconsistent state update vulnerabilities can be classified into four categories, as shown in Table~\ref{exposure}.

\begin{table}
\caption{Category of the exploitation methods for inconsistent state update vulnerabilities.}
\begin{center}
\begin{tabular}{ l c c }
\toprule
\textbf{Exploitation method categories} & \textbf{Instance} & \textbf{Percentage}  \\
\midrule
Exploiting Numerical Calculation Errors & 65 & 56.03\% \\
Repeated Transactions & 27 & 23.28\% \\
Interim State Exploits & 15 & 12.93\% \\
Paralyzing Contract Functionality & 9 & 7.76\% \\
\bottomrule
\end{tabular}
\label{exposure}
\end{center}
\end{table}

\subsubsection{Exploiting Numerical Calculation Errors (56.03\%).}
During the normal operation of a contract, critical computational processes (such as reward distribution, asset liquidation, and share profit calculations) may experience minor deviations in their results due to {inconsistently updated states}. Attackers actively seek and exploit these discrepancies to amplify their impact and gain unfair advantages. 
The exploitation typically requires analyzing the contract code to identify potential opportunities. For instance, attackers may first locate code segments susceptible to calculation errors due to state inconsistencies and determine the potential for profit.
Then, based on data flow relationships, they identify transactions related to these calculation errors. Finally, they call functions that can enable them to illegally profit, thereby obtaining illicit gains.

For instance, during asset liquidation, if interest rates are not accumulated correctly, attackers can bypass the required interest by performing self-liquidation. Similarly, in cases where the contract fails to properly calculate account assets after a transaction, a discrepancy may arise between the user's assets and the actual situation, allowing attackers to withdraw excess funds. Additionally, miscalculations in reward distribution can lead to reward discrepancies, enabling attackers to fabricate higher distribution amounts and thus obtain higher returns in the reward calculation. Beyond these, in an on-chain, transparent, and multi-contract interaction environment, attackers can also exploit partial failures in batch operations or errors in the states passed between contracts to further disrupt computational consistency and profit from these flaws. When these vulnerabilities are successfully exploited, they typically result in direct economic losses to users or protocol pools, threatening the security and economic stability of the system. 
To alleviate the issue, contract coding should actively adopt effective defense
mechanisms to block numerical calculation errors at much as possible and three potential solutions arise. First, the  \texttt{require()} function can be used to validate transaction preconditions. Second, access control and state encapsulation can be leveraged  to restrict unintended modifications of critical states. Finally, the \texttt{assert()} function can be applied after core computation steps to verify whether the results comply with inherent constraints of transaction logic.

We next illustrate this exploitation method through how the vulnerability in Fig.~\ref{fig:Computation} is exploited,
and Fig.~\ref{fig:Computation_flow} visualizes the exploitation flow. First, the attackers wait for or actively force the variable \texttt{startSupply} to become zero (for example through large-scale redemption operations). At this point, the \texttt{lastFee} timestamp becomes fixed as T0. After some time, an unsuspecting victim performs the first mint operation at time T1. Since \texttt{startSupply} equates to 0, \texttt{handleFees()} returns immediately without updating \texttt{lastFee}. Then, at time T2 (just seconds after T1), the attackers immediately execute a second mint operation. The contract now erroneously calculates and mints a large number of tokens based on the excessive time difference between T0 and T2 (when only the few seconds between T1 and T2 should actually be calculated). These additionally minted tokens dilute the victims' asset value, while the attackers obtain these excess minted tokens.
\begin{figure}[htbp]
\centering
\includegraphics[width=0.7\textwidth]{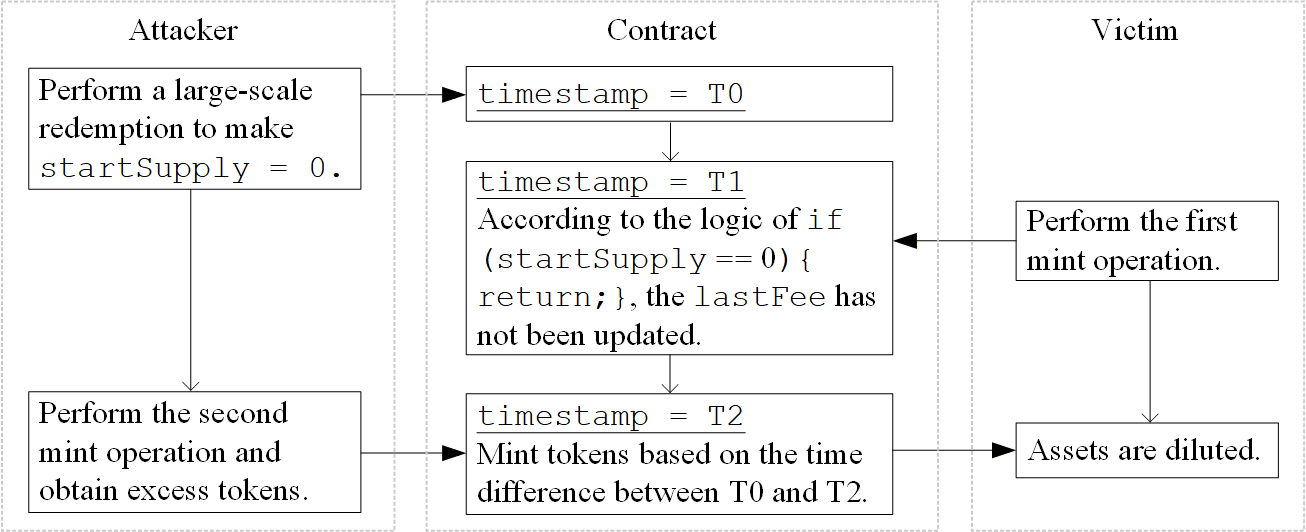}
\caption{Exploitation flow for the vulnerability in the time-based fee asset vault contract.}
\label{fig:Computation_flow}
\end{figure}

\vspace{1mm}
\begin{center}
\fcolorbox{black}{gray!25}{\parbox{0.97\linewidth}{
\textit{\noindent\textbf{Finding 8}:
Among all attacks on top of inconsistent state update vulnerabilities, around sixty percentage of them exploit numerical calculation errors, meaning that attackers seek and exploit computational discrepancies to amplify their impact and gain unfair advantages.
}

\textit{
\noindent\textbf{Implication 8}: To mitigate risks from numerical calculation errors, contract coding should actively adopt effective defense mechanisms. Potential solutions include: (1) using the Solidity {require()} function to validate transaction preconditions, (2) leveraging access control and state encapsulation to restrict unintended modifications of critical states, (3) applying the Solidity assert() function after core computation steps to verify whether the results comply with the inherent constraints of transaction logic. 
}}}
\end{center}
\vspace{1mm}

\subsubsection{Repeated Transactions (23.28\%).}
This exploitation method manifests when an attacker repeatedly performs the same operation to exploit the inconsistently updated states for undue profit. More specifically, attackers may input identical conditions (such as the same token address or order) or invoke functions like reward claiming and asset transfer multiple times, leading to significant economic losses.
Such actions may ultimately deplete liquidity pools, inflate token supply, or result in theft of user assets, thereby compromising the logical integrity of the contract. Since these attacks often involve legitimate function calls, relying solely on input parameter checks is typically insufficient for effective detection. This type of attack is characterized by repetitive actions, incurs low cost, and does not require complex inter-contract interactions. Exploiting a single logical flaw can significantly amplify profit and even drain the contract’s funds. To prevent such issues, developers must ensure that state variables are updated accurately and promptly after each operation. An effective mitigation is to incorporate unique identifiers—such as random numbers or counters—to prevent repeated executions. In addition, for any code logic that may involve multiple repeated actions, setting sufficient cooldown periods can effectively reduce the risks caused by consecutive operations. In particular, repeated reward distributions and share adjustments should restrict the same user from performing identical actions repeatedly within a short timeframe.

We next illustrate this exploitation method through how the vulnerability in 
Fig.~\ref{fig:Initialization} is exploited, and Fig.~\ref{fig:Initialization_flow} visualizes the exploitation flow.
For this vulnerability, attackers first call function \texttt{initLPP()} to create a proposal. 
Note that the \texttt{initLPP()} function is used to set the metadata of a proposal.
After the proposal passes the challenge period, they call the function again, using \texttt{setPartOffset(\_partOffset).setClaimedSize()} to overwrite the original \texttt{metaData} in storage. The \texttt{proposals.push()} operation redundantly adds the same key, while the bond amount in \texttt{proposalBonds[msg.sender]} is reset. These operations maliciously tamper with the original proposal's metadata, causing the contract to process it incorrectly in subsequent steps. As a result, the attackers can exploit the dispute mechanism to fraudulently win claims.
\begin{figure}[htbp]
\centering
\includegraphics[width=0.68\textwidth]{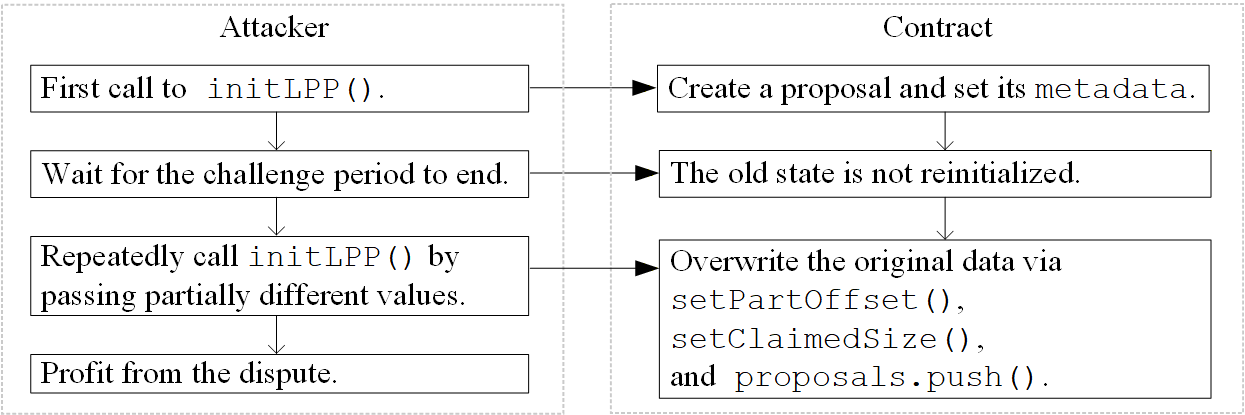}
\caption{Exploitation flow for the vulnerability in the proposal validation and dispute resolution contract.}
\label{fig:Initialization_flow}
\end{figure}

\vspace{1mm}
\begin{center}
\fcolorbox{black}{gray!25}{\parbox{0.97\linewidth}{
\textit{\noindent\textbf{Finding 9}: 
More than twenty percentage of exploitations for inconsistent state update vulnerabilities make use of repeated transactions, where an attacker
triggers operations multiple times within a short period and exploits stale state data to obtain repeated profits.
}

\textit{
\noindent\textbf{Implication 9}: Defensive designs can incorporate unique identifiers to prevent repeated execution, e.g., through random numbers or counters. Besides, for any logic that may involve multiple repeated actions, setting sufficient cooldown periods can effectively reduce the risks caused by consecutive operations. In particular, repeated reward distributions and share adjustments should restrict the same user from performing identical actions repeatedly within a short timeframe.
}}}
\end{center}
\vspace{1mm}

\subsubsection{Interim State Exploits (12.93\%).}
This exploitation method occurs during the interim phase of a transaction, when critical state updates have not yet been completed. Attackers exploit this brief window of delayed state synchronization to insert malicious operations and extract undue profit. Typical manifestations of this exploitation method include price manipulation, front-running, and reentrancy attacks.

Price manipulation often arises when the contract relies on outdated historical prices or unsynchronized oracle data. In particular, historical prices can influence commodity pricing by affecting calculation mechanisms such as Time-Weighted Average Price (TWAP). Attackers may execute arbitrage trades before the oracle reflects the actual market price, or even tamper with the oracle source itself, thereby distorting asset valuations to siphon funds or manipulate market dynamics. To reduce the risk of price manipulation, developers should ensure that the data update frequency of oracles or TWAP aligns with the contract’s price reading cadence, minimizing or eliminating exploitable delay windows. Potential defensive strategies include employing multiple oracle mechanisms, updating prices prior to liquidity operations, and implementing operation cooldown periods.
Examples in Fig.~\ref{fig:Price} and Fig.~\ref{fig:Incorrect-Logic} illustrate vulnerabilities that enable price manipulation through inconsistent state updates.
For the vulnerability in Fig.~\ref{fig:Price}, its exploitation flow is shown in Fig.~\ref{fig:Price_flow}. More specifically, attackers can first sell large amounts of VADER tokens to drive down the market price, and then immediately call the \texttt{\_updateVaderPrice()} function to synchronize prices. Since this function doesn't update the historical VADER price in the \texttt{previousPrices} array, the subsequent \texttt{currentLiquidityEvaluation} calculation mistakenly uses the initial higher price to assess VADER's value (while the actual market price has been lowered). The attackers then exchange other assets for the overvalued VADER at a low price, and complete the arbitrage by selling them after the market price recovers.
For the vulnerability in Fig.~\ref{fig:Incorrect-Logic}, its exploitation flow is shown in Fig.~\ref{fig:Incorrect-Logic_flow}. In particular, the malicious lenders call the \texttt{updateLoanParams()} function and pass a \texttt{params} struct containing manipulated oracle addresses. When the collateral's market value increases, the attackers invoke the \texttt{removeCollateral()} function to obtain collateral valuations. Since the provided oracles are controlled by the attackers, they can deliberately report valuations significantly lower than the actual market prices. This causes the ``\texttt{rate.mul(loanParams.ltvBPS)/BPS $<$ amount}'' check condition to be satisfied, thereby triggering the collateral seizure logic. Ultimately, the attackers illegally obtain collateral whose real value far exceeds the principal and interest of the loans.

\begin{figure}[htbp]
\centering
\includegraphics[width=0.85\textwidth]{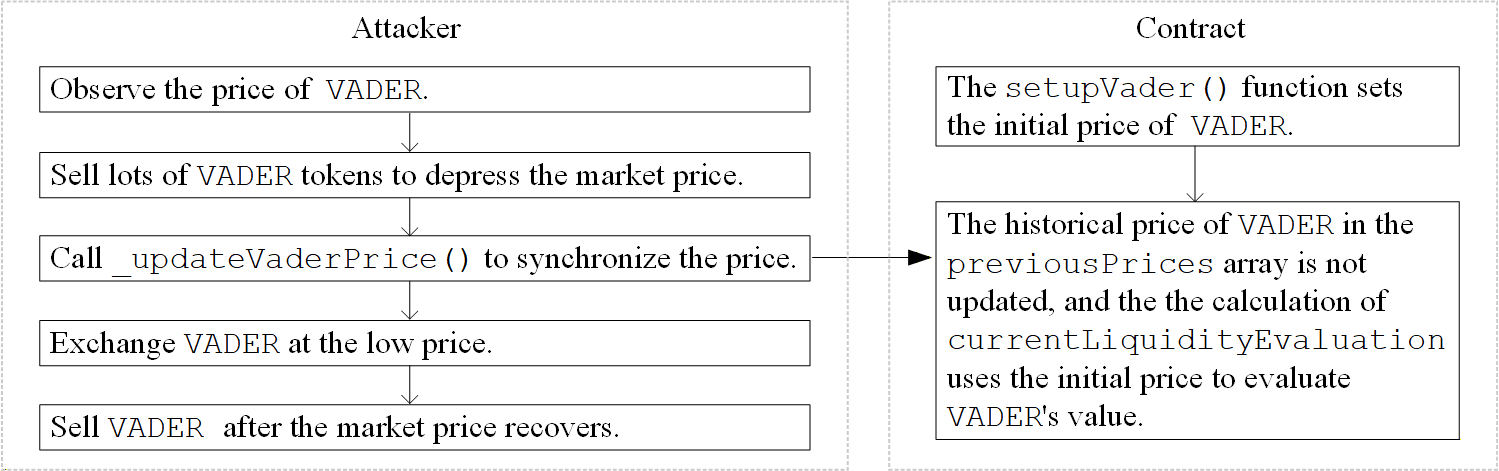}
\caption{Exploitation flow for the vulnerability in the liquidity-weighted price oracle contract.}
\label{fig:Price_flow}
\end{figure}

\begin{figure}[htbp]
\centering
\includegraphics[width=0.88\textwidth]{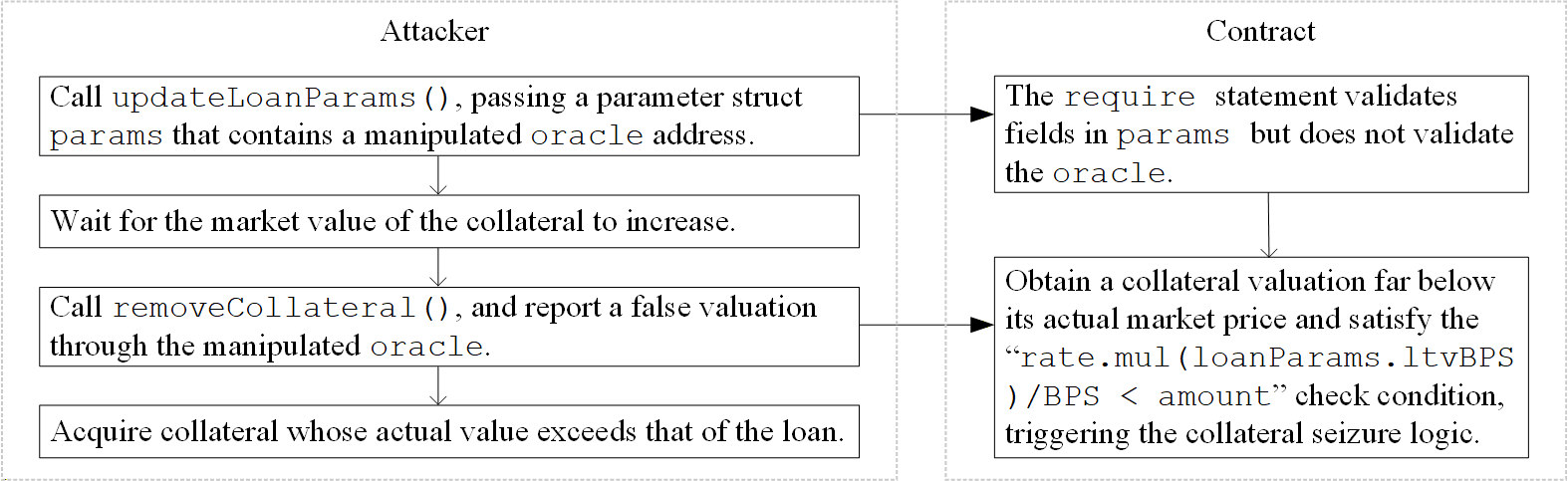}
\caption{Exploitation flow for the vulnerability in the oracle-driven asset-backed loan contract.}
\label{fig:Incorrect-Logic_flow}
\end{figure}



Front-running involves preemptively inserting transactions to interfere with the intended state transition path, thereby gaining profit. This commonly occurs in scenarios where slippage protection is absent or reward mechanisms are not properly reset. Attackers may withdraw assets before new limits take effect, and may also combine flash loans with dynamic parameter adjustment mechanisms to create multi-layered interference at the block level, thereby amplifying the attack. 
To alleviate the issue, introducing operation ordering locks or uniqueness flags are beneficial as they can ensure that transactions are executed only according to the intended flow. Additionally, implementing economic mechanisms like operation deposits or penalties can increase the attack cost, making the attack economically unprofitable.
For the vulnerability in Fig.~\ref{fig:Front-running}, its exploitation flow is given  in Fig.~\ref{fig:Front-running_flow} and the attacker may unlawfully increase the loan amount before the loan is issued. Here, the lender signs a loan agreement and intends to optimize existing loan terms through the \texttt{refinanceFull()} function. An attacker monitoring the transaction mempool obtains the signature and calls the \texttt{addNewTranche()} function before the function \texttt{refinanceFull()} executes, submitting the same agreement data and signature as the lender. Since the function \texttt{addNewTranche()} only verifies signature validity without checking the agreement's intended purpose, the operation passes validation. The contract mistakenly interprets the signature - meant for loan optimization - as authorization to add a new loan tranche. This creates an additional loan tranche on top of the original loan, doubling the total loan amount and forcing the lender to bear significantly higher risk than intended.

\begin{figure}[htbp]
\centering
\includegraphics[width=0.97\textwidth]{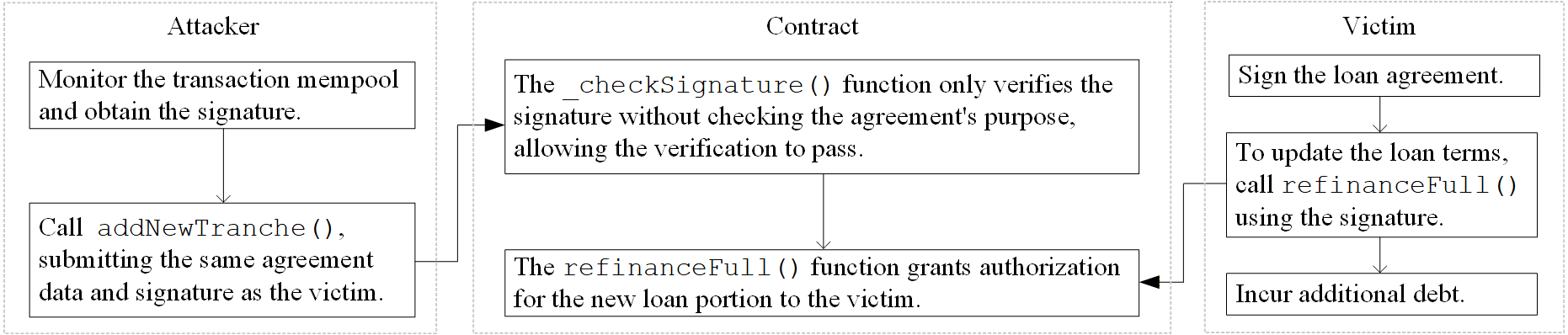}
\caption{Exploitation flow for the vulnerability in the flexible loan contract.}
\label{fig:Front-running_flow}
\end{figure}

For reentrancy attacks, the attackers deploy a malicious contract with a callback mechanism to re-enter the target function before the function completes its state update, allowing them to invoke sensitive logic such as claiming restricted assets or manipulating share calculations. 
For the vulnerability in Fig.~\ref{fig:Reentrancy}, its exploitation flow is shown in Fig.~\ref{fig:Reentrancy_flow} and the attacker performs a reentrancy by exploiting the state variable \texttt{commandId} that was not updated in time. Attackers first initiate a transfer, then immediately call the \texttt{expressReceiveTokenWithData()} function before the target chain transaction executes. Within this function, \texttt{safeTransferFrom()} transfers tokens from the caller's account to the attackers' malicious contract address \texttt{destinationAddress}. At this point, the malicious contract is triggered and it calls back the original transfer. Since the function \texttt{\_setExpressReceiveTokenWithData()} has not been executed yet, the transaction status remains uncompleted, allowing the original transfer to be executed again. As a result, the attackers' malicious contract receives double tokens.

\begin{figure}[htbp]
\centering
\includegraphics[width=0.93\textwidth]{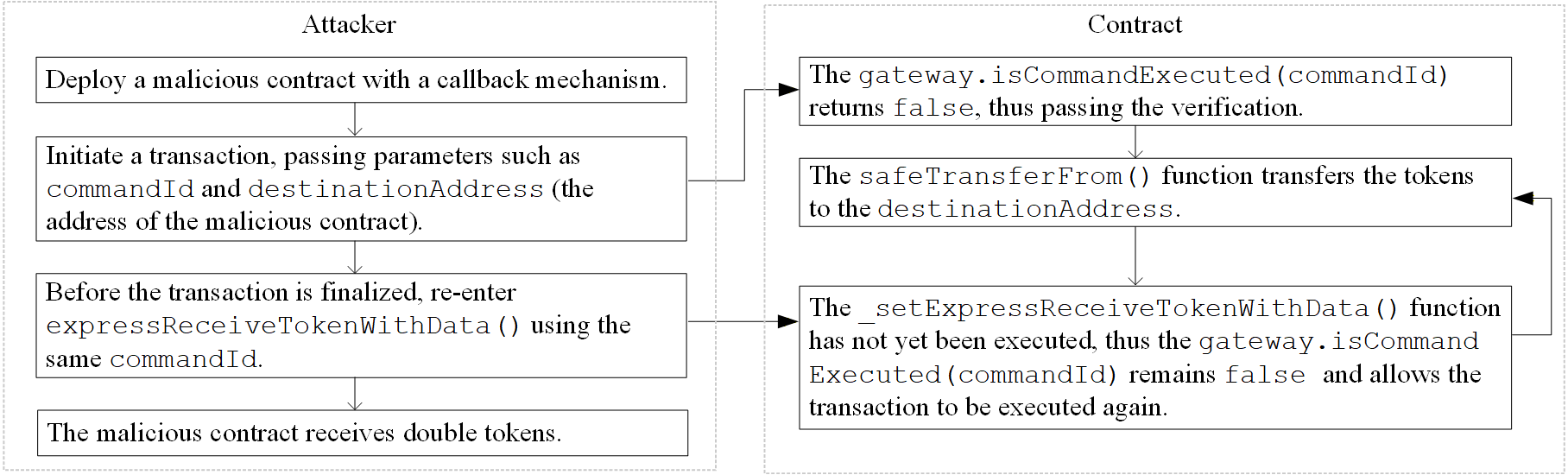}
\caption{Exploitation flow for the vulnerability in the interchain token transfer service contract.}
\label{fig:Reentrancy_flow}
\end{figure}

The core danger of interim state exploitation lies in attackers precisely hijacking a single moment during contract execution to steal funds or distort contract logic. This type of attack does not aim to disrupt the contract, but seeks illicit gains within the gaps of its normal operation. By manipulating outdated or unsynchronized price data, attackers can directly engage in arbitrage, illegally transferring assets far exceeding their appropriate value. Through front-running, attackers can hijack legitimate operational intents and maliciously alter key protocol parameters such as loan limits. Via reentrancy attacks, attackers can repeatedly trigger fund transfer logic before state updates are completed, resulting in multiplied asset theft. Table~\ref{InterimTable} gives a summary of the three interim states exploits. Overall, by conspiratorially using the normal business processes of a contract, interim state exploitation depletes the protocol's value and subverts its intended operational rules in a highly covert and technical manner.
Developers should adopt strategies such as prioritizing state updates, restricting execution paths, and implementing slippage protection to minimize the exposure window and prevent the contract from being maliciously manipulated during this intermediate state.

\begin{table}
\caption{A summary of the three interim states exploits.}
\begin{center}
\begin{tabular}{ l l l l}
\toprule
\textbf{\makecell[l]{Attack types of \\interim state exploits}} & \textbf{Exploited interim state} & \textbf{Primary impact} & \textbf{Defense strategies} \\
\midrule
Price manipulation & \makecell[l]{The delay window between\\ market price updates and \\the recorded price within \\the contract.} & \makecell[l]{Leads to arbitrage due\\ to distorted asset \\valuation.} & \makecell[l]{Multi-oracle price \\verification with timely \\on-chain synchronization.} \\
\midrule
Front-running & \makecell[l]{The time window after a \\transaction intent is\\ revealed but before the \\on-chain state changes.} & \makecell[l]{Results in loss of profits \\through the hijacking\\ of legitimate operations.} &\makecell[l]{Introduce operation \\ordering locks or \\uniqueness flags.} \\
\midrule
Reentrancy attack & \makecell[l]{The brief window between \\the actual outflow of\\ funds and the update of\\ the ledger balance.} & \makecell[l]{The transfer logic was\\ triggered multiple times \\and the funds were \\stolen.} & \makecell[l]{Adhere to the \\Checks-Effects-Interactions \\pattern, and use reentrancy \\guard modifiers.} \\
\bottomrule
\end{tabular}
\label{InterimTable}
\end{center}
\end{table}

\begin{center}
\fcolorbox{black}{gray!25}{\parbox{0.97\linewidth}{
\textit{\noindent\textbf{Finding 10}: 
Around fifteen percentage of exploitations for inconsistent state update vulnerabilities belong to interim state exploits, and three typical manifestations include price
manipulation, front-running, and reentrancy attacks. Price
manipulation arises if there is a delay window between data updates and price calculation when using on-chain oracles or TWAP for pricing, and front-running typically profits from seizing specific operation timings. 
}

\textit{
\noindent\textbf{Implication 10}: 
To reduce interim state exploits,  defensive designs should minimize the
exposure window and prevent the contract from being maliciously manipulated during this intermediate state. In particular, to reduce the risk of price manipulation, possible defensive strategies include employing multiple oracle mechanisms, updating prices prior to liquidity operations, and implementing operation cooldown periods. To reduce the risk of front-running, introducing operation ordering locks or uniqueness flags are beneficial as they can ensure that transactions are executed only according to the intended flow.
}}}
\end{center}
\vspace{1mm}

\subsubsection{Paralyzing Contract Functionality (7.76\%). }
Attacks on top of this exploitation method are characterized by disrupting the normal logic of the contract, preventing its key functions from operating properly. Attackers take advantage of flaws where critical state variables are not updated correctly after specific operations (\emph{e.g.}, updating management addresses, cross-chain message processing, \emph{etc.}), and use abnormal operation sequences (which overwrite key state variables) or input extreme parameters to disrupt the management of key states. This leads to the failure of subsequent modules that depend on these state variables (\emph{e.g.}, permission validation, message forwarding decisions, \emph{etc.}) or results in permanent asset lockup. For instance, an attacker may submit a new proposal while an old proposal is still being executed, overwriting the hash of the old proposal and causing a failure in fund release verification. 
Additionally, in cross-chain message processing, an attacker may inject an empty string to overwrite a user-specified message field, causing the target chain's contract to fail in receiving a valid message. The specific impacts of such vulnerabilities include, but are not limited to, disruption of contract logic, conditions permanently failing to meet requirements, and system stagnation. This not only compromises the integrity of the contract's business logic but also poses a direct threat to user trust and the security of on-chain assets. To effectively alleviate the issue of paralyzing contract functionality, potential efforts can be made in two directions. First, to defend against state overwrite attacks, critical operations can first copy global state variables into local variables, ensuring that subsequent logic relies solely on the cached values. This prevents external calls from overwriting the original variables and breaking the intended logic. Second, to resist extreme input attacks,
sensitive operations (such as large withdrawals) can be restricted by setting per-operation thresholds, avoiding functional failures caused by abnormal inputs.


Fig.~\ref{fig:FunctionFailure} presents an example vulnerability to illustrate this exploitation method, and its exploitation flow is shown in Fig.~\ref{fig:FunctionFailure_flow}. 
The corresponding contract is used to manage NFT orders, including creating sales orders, verifying whether the orders are completed, and distributing the proceeds after the sale is completed. The \texttt{\_constructOrder()} function is responsible for creating a new order. It generates an order hash based on the input parameters and stores it in \texttt{vaultOrderHash[\_vault]}, thereby overwriting the previous hash. When a user later attempts to withdraw the sales proceeds via the \texttt{cash()} function, the contract calls the \texttt{\_verifySale()} function to verify the order and uses the current \texttt{vaultOrderHash[\_vault]} to check its fulfillment status. 
If an order that was previously sold has not been fully processed, attackers can propose a new proposal and overwrite the old proposal's hash with the new proposal's hash, causing the verification to fail. This ultimately results in the failure of the cash() function, and the sale proceeds are permanently locked in the contract.
To prevent such withdrawal failures caused by overwritten order hashes, the hash should instead be stored in \texttt{activeListings[\_vault]}, which records the currently valid listing information. The \texttt{cash()} function should then verify the order status based on this stored hash.

\begin{figure}[htbp]
\centering
\includegraphics[width=0.8\textwidth]{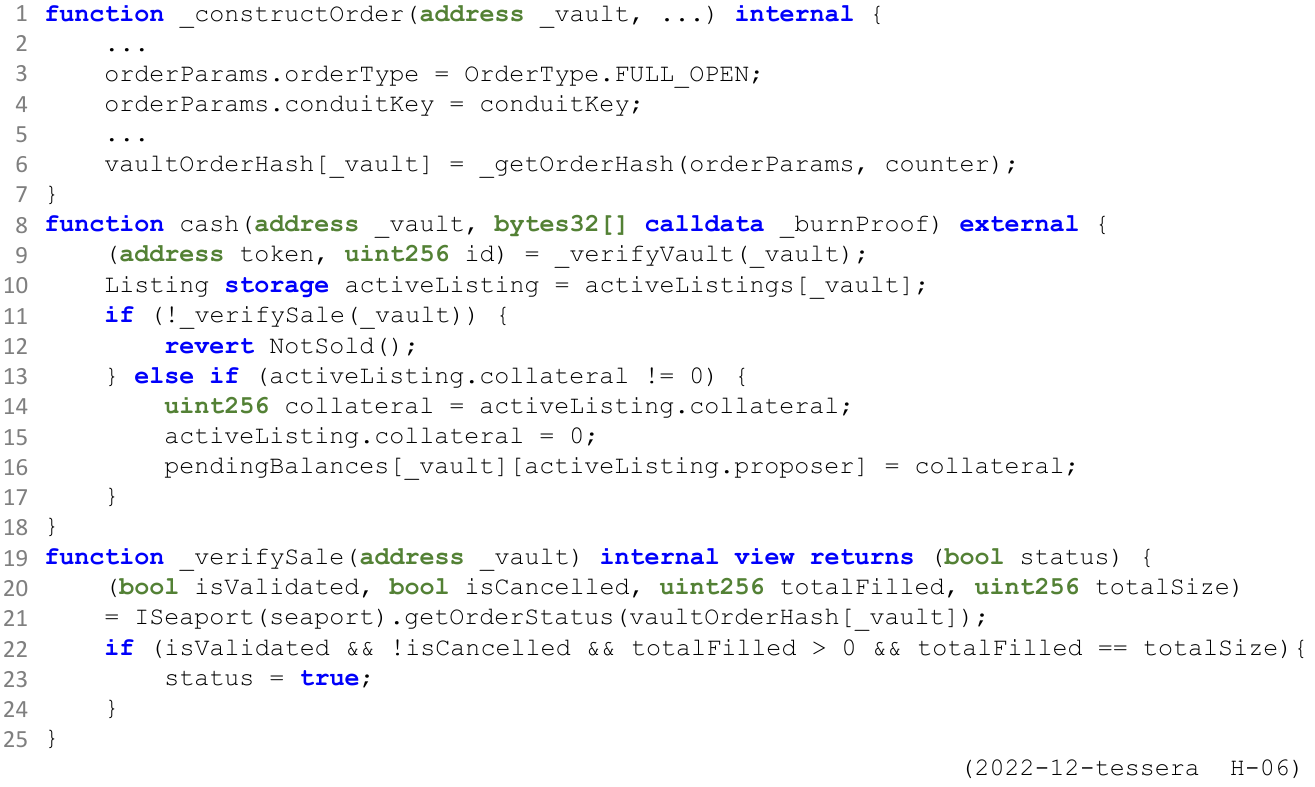}
\caption{A vulnerability in an order management contract (\url{https://code4rena.com/reports/2022-12-tessera}).}
\Description{A function failure bug.}
\label{fig:FunctionFailure}
\end{figure}

\begin{figure}[htbp]
\centering
\includegraphics[width=0.94\textwidth]{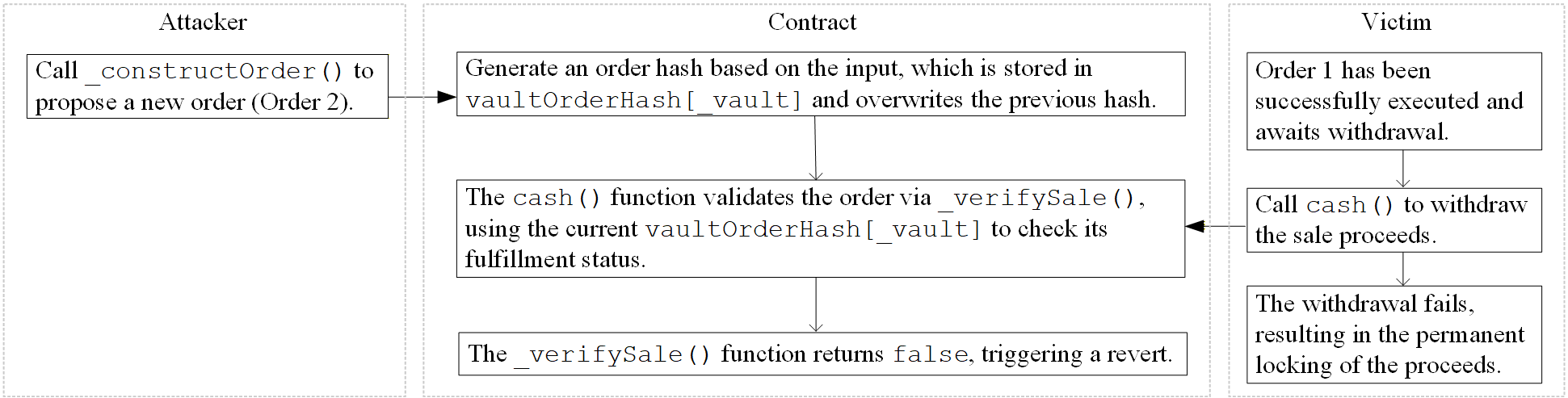}
\caption{Exploitation flow for the vulnerability in the order management contract.}
\label{fig:FunctionFailure_flow}
\end{figure}

\begin{center}
\fcolorbox{black}{gray!25}{\parbox{0.97\linewidth}{
\textit{\noindent\textbf{Finding 11}: 
Around eight percentage of exploitations for inconsistent state update vulnerabilities belong to paralyzing contract functionality, which generally leads to irreversible failure of the contract’s core functions. Two typical manifestations are state overwrite attacks and extreme input attacks.
}

\textit{
\noindent\textbf{Implication 11}: 
To alleviate the issue of paralyzing contract functionality, potential efforts can be made in three
directions. First, to defend against state overwrite attacks, critical operations can first copy global state variables into local variables, ensuring that subsequent logic relies solely on the cached values. 
Second, to resist extreme input attacks,
sensitive operations (such as large withdrawals) can be restricted by setting per-operation thresholds.
Finally, adopting the proxy contract development pattern enables developers to restore project functionality after a contract becomes paralyzed, thereby preventing the permanent locking of user assets.}}
}
\end{center}
\vspace{1mm}

\subsection{Correlation between the Root Causes and Exploitation Methods}

\begin{figure}[htbp]
\centering
\includegraphics[width=0.80\textwidth]{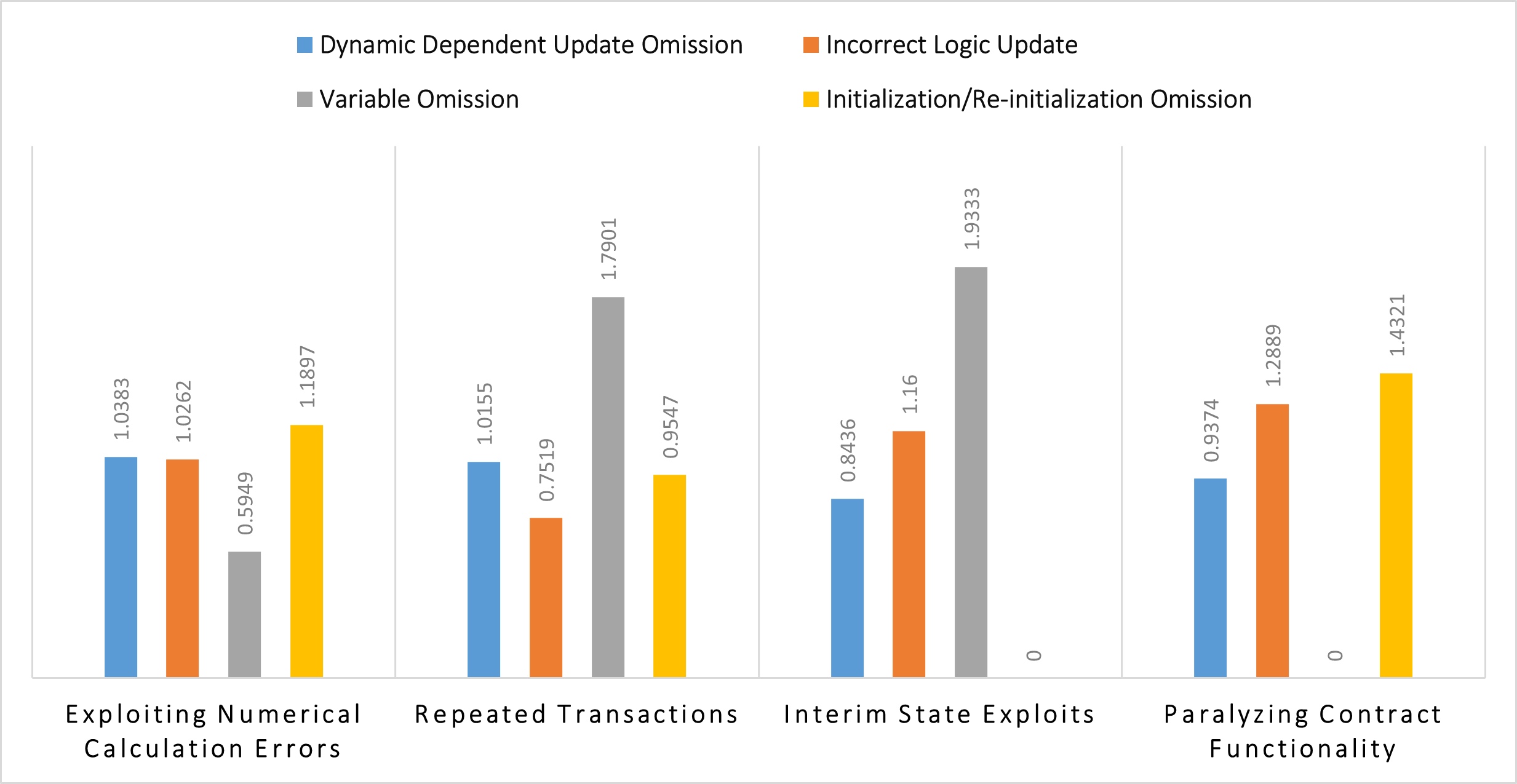}
\caption{The statistical metric \emph{lift} between the root cause of an inconsistent state update vulnerability and its exploitation method.}
\Description{The statistical metric \emph{lift} between the root cause of an inconsistent state update bug and its exploitation method.}
\label{lift_Exploitation}
\end{figure}

To understand the relationship between the cause of an inconsistent update vulnerability and its exploitation method, we calculate their \emph{lift}, following the approach in Sec.~\ref{Correlation between Root Causes and Fix Strategies}. If the calculated value \emph{lift} is equal to 1, then the root cause \emph{A} is independent with the exploitation method \emph{B}. If the calculated \emph{lift} value is larger than 1, then the root cause \emph{A} and the exploitation method \emph{B} are positively correlated. The positive correlation means that if an inconsistent state update vulnerability is caused by \emph{A}, then it is more likely to be exploited through method \emph{B}. If the calculated \emph{lift} value is smaller than 1, then \emph{A} and \emph{B} are negatively correlated.

The result is given in Fig.~\ref{lift_Exploitation}. From this figure, we can see that 1) for the root cause category ``Dynamic Dependent Update Omission'', it has positive correlations with exploitation method categories ``Exploiting Numerical Calculation Errors'' and ``Repeated Transactions”; 2) for the root cause category ``Incorrect Logic Update'', it has positive correlations with exploitation method categories `` Exploiting Numerical Calculation Errors'', ``Interim State Exploits” and ``Paralyzing Contract Functionality''; 3) for the root cause category ``Variable Omission'', it has positive correlations with exploitation method categories ``Interim State Exploits'' and ``Repeated Transactions''; 4) for the root cause category ``Initialization/Re-initialization Omission'', it has positive correlations with exploitation method categories ``Paralyzing Contract Functionality'' and ``Exploiting Numerical Calculation Errors''. 
Overall, compared with the high correlation between vulnerability causes and fix strategies, the correlation between vulnerability causes and exploitation methods is relatively weak.


\section{Checker for State Variable Update Omission and State Variable Optimization}
\label{checker}
To demonstrate the potential value of our findings, we design a checker to detect a specific type of problematic code, which may involve missing state variable updates or gas consumption that can be optimized. The checker is initially motivated by the large percentage of inconsistent state update vulnerabilities that arise due to the omission of update for dynamic operation dependent state variables (Finding 1).

\subsection{Design and Implementation of the Checker}
Our insight is that when a developer declares a state variable in a normal smart contract, if the declaration does not use the \texttt{constant} or \texttt{immutable} modifier and the variable is never reassigned throughout the project, then it can be concluded that the update for that variable has been missed—assuming that the developer did not forget to use these two modifiers (\emph{i.e.}, \texttt{constant} or \texttt{immutable}) during declaration. 
Therefore, when the checker detects a state variable that meets the above conditions, it will issue a warning, indicating that the variable may have a missing update or may require the use of the modifier \texttt{constant} or \texttt{immutable} to improve the contract's security and optimize gas consumption.

For state variables with special purposes that do not require future changes, such as maximum supply, contract administrator addresses, and configuration information, we remind developers to declare them with the \texttt{constant} or \texttt{immutable} modifier as needed. State variables declared with these two modifiers do not occupy storage space in the EVM, as their values are directly embedded into the contract’s bytecode during compilation. This approach effectively eliminates storage and read operations on these variables within the EVM, thereby avoiding the additional overhead of SLOAD (2,100 gas for the first read per transaction, 100 gas for each subsequent read) and SSTORE (minimum 20,000 gas).

The checker is built upon \emph{solidity-parser} \cite{solidity-parser}, which is capable of parsing multiple versions of the Solidity language and generating an Abstract Syntax Tree (AST).  The checker relies on the parsed AST to function. As shown in Fig.~\ref{fig:Checker}, it takes the project directory path as input and then analyzes all the declared state variables throughout the project. The specific checks consist of two main stages. First, it identifies all the state variables in the project's contracts that are not declared with the \texttt{constant} or \texttt{immutable} modifier. Then, for each identified state variable, it checks whether it is reassigned anywhere in the project. If the variable is not reassigned, it is considered problematic. The output of the checker includes the following information: (1) the problematic state variable, and (2) the directory of the file and the detailed location where the state variable is declared.

\begin{figure}[htbp]
\centering
\includegraphics[width=0.58\textwidth]{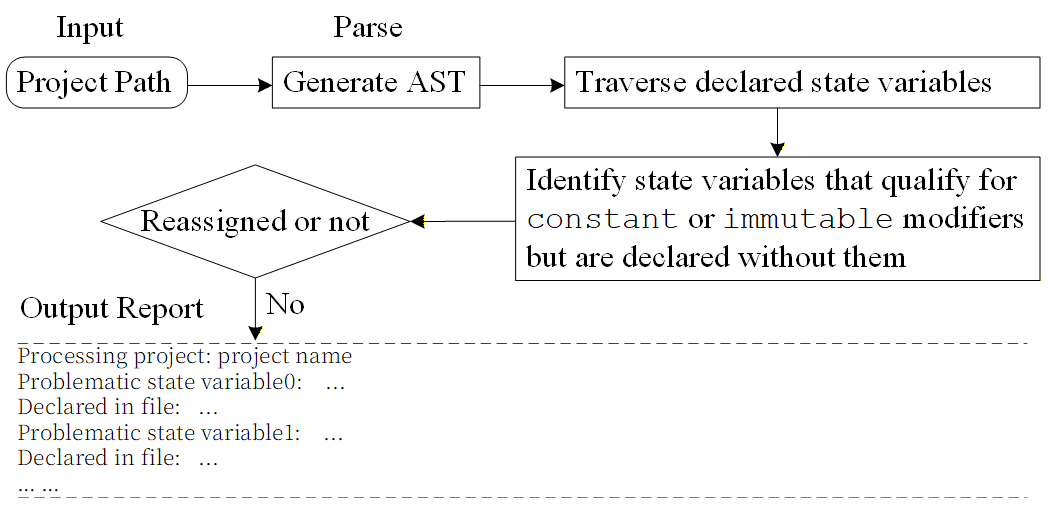}
\caption{Workflow of the checker for state variable update omission and state variable optimization.}
\label{fig:Checker}
\end{figure}

\subsection{Evaluation of the Checker}
\label{checkerEvaluation}
To investigate the effectiveness of our tool on popular projects, we applied the checker to 208 popular and active Solidity projects on GitHub. In particular, the 208 projects have been updated within the past month (from February 8 to March 8, 2025) with respect to our evaluation time to ensure their activeness.
They also have at least 30 stars to ensure the popularity, since the number of stars has been widely used as a proxy for the popularity of GitHub projects \cite{xiao2023early, borges2018s}.
In addition, to minimize the inclusion of Solidity code used for testing or deployment \cite{foundry,hardhat}, we removed all directories and files named \emph{test}, \emph{script}, \emph{mock}, and \emph{deploy} before conducting the local evaluation.

The checker identified the targeted issue in the latest code versions of 64 projects. We then submitted issues for each affected project, including the following information: (1) a concise message explaining the purpose of our study; (2) an explanation of our findings regarding state variable update omissions and state variable initialization optimizations; (3) the exact state variables for which updates were omitted or could be optimized based on our findings (including the associated code snippets). 
Within two weeks of submitting the issues, the owners of 23 projects responded. The types and quantities of the received responses are  shown in Table~\ref{evaluation}.

Among the 23 responses, 19 confirmed the effectiveness of our tool. General issues identified by the checker include missed state updates and failure to use the appropriate \texttt{constant} or \texttt{immutable} modifiers. Additionally, several special cases were mentioned in the responses: (1) developers refuse to use gas-saving modifiers due to a lack of concern for gas costs (1 response); (2) key state updates are located in private code (1 response); (3) 
the problematic code is either considered not worth optimizing due to its limited importance to their libraries (1 response); (4) the problematic code is intended solely for example demonstration (2 responses); and (5) deliberate vulnerabilities that are used as fuzzing targets (1 responses). Overall, these special cases account for 1 instance of state update omission and 5 instances of forgetting to use specified modifiers, respectively. In actual deployments, these
special cases do not occur, and thus identifying predefined problematic code under these conditions remains strong evidence of our tool's effectiveness.
Our work has received positive feedback from many Solidity developers, with some describing our research as ``meaningful'' and others calling it a ``good shout!''. These results indicate that, beyond our specific findings, this proof-of-concept checker can effectively enhance the security of state variable updates in Solidity smart contracts.

\begin{table}
\caption{Category of the developer responses for the checker.}
\begin{center}
\begin{tabular}{ l c c }
\toprule
\textbf{Response categories} & \textbf{Instance} \\
\midrule
\makecell[l]{State update omission} & 6 \\
Forgetting to use specified modifiers & 13  \\
False positive & 4 \\
\bottomrule
\end{tabular}
\label{evaluation}
\end{center}
\end{table}


Besides the 19 confirmed issues, there are also 4 false positives in our detection results. Note that if the checker we develop is perfect, there will be no false positives or missed detections. Due to the imperfection of the checker in its current form, false positives or missed detections may occur. 
For the 4 false positives, there exist two situations.
One situation is although we have removed the test directories according to Solidity development practices before the detection, the inspected projects still inevitably contain test-related files due to varying coding habits among developers (2 responses). Another situation involves retaining the storage layout during upgrades to ensure backward compatibility (2 responses).
These false positives suggest that our tool still requires further improvement to eliminate such cases.
To address the aforementioned false positives, the next improvements for our tool will focus on deepening its semantic analysis and context-aware capabilities. On the one hand, we will build a comprehensive pattern list for tests and combine it with function call graph analysis to accurately identify and filter scattered test code and their associated state variables. On the other hand, we will enhance the recognition of upgradeable contract systems by identifying storage gap variables (\emph{i.e.,} intentionally declared but unused state variables reserved for future upgrades) and proxy pattern contexts. In this way, we can effectively distinguish truly problematic variables from those retained for storage compatibility, thus significantly improving detection accuracy. Besides, we will also work towards addressing potential false negatives to improve the completeness of detection. Some projects suffer from incomplete dependency information that causes certain compilation failures, thus the checker cannot acquire declarations of certain state variables and produce false negatives. We plan to enhance the checker’s parsing and fault-tolerance capabilities so as to identify and cover all variable declarations as completely as possible across various scenarios.

\section{Discussion}

Our study reveals that nearly half (47.41\%) of inconsistent state update vulnerabilities originate from ``Dynamic Dependent Update Omission''. This finding not only confirms that this issue is a pervasive pain point in smart contract development, but also exposes potential limitations in existing formal verification and static analysis tools. While these tools excel at detecting implemented code, they often prove ineffective against update operations that should exist but were omitted by developers. This insight directs research attention toward novel detection paradigms with deeper semantic understanding.
Furthermore, our study also reveals a strong correlation between the root causes of vulnerabilities and their fix strategies. 
This high degree of association indicates that inconsistent state update vulnerabilities caused by specific root causes are more likely to be fixed through their corresponding specific strategies. Establishing a probabilistic mapping from root causes to fix strategies can provide a reliable foundation for building rule-based automated repair systems.

Integrating the implications from all of our findings, we provide consolidated recommendations for different stakeholders. For developers and practitioners, the core implication is elevating state synchronization to a primary design principle. This involves clearly identifying related state variables during the design phase, strictly adhering to the CEI pattern to defend against interim state exploits, and adopting defensive programming techniques to mitigate risks like numerical calculation errors and repeated transactions. For researchers and tool developers, our study calls for moving beyond syntactic pattern matching toward techniques capable of understanding data flow and semantic context. In addition, the clear taxonomy and dataset we provide also offer a valuable foundation for exploring AI-driven patch generation and extending research to other blockchain platforms.

Within the broader research context, this study fills a gap in the smart contract security field. Previous studies mainly focused on vulnerabilities with relatively obvious code patterns, but the more concealed issue of inconsistent state updates has rarely been explored. By providing fine-grained classification, real-world cases, and quantitative data, this study establishes a strong foundation for future related research. Our study also proves that ensuring the consistency of state updates is a fundamental and severe challenge, and we have provided a way for building a more secure and reliable smart contract ecosystem through empirical understanding and tool-driven solutions.

\section{Threats to Validity}
The \emph{internal validity} is primarily related with the manual process that can possibly introduce errors. To mitigate this threat, following a mature and widely used process, we ensure that each vulnerability has been examined by at least 
two experienced Solidty smart contract developers and researchers.
They thoroughly analyze the source code, code comments, bug reports, and fix suggestions of the relevant contracts to  understand the vulnerabilities. In addition, the whole artifact for this paper is available online to let readers gain a more deep understanding of our study and analysis. 

The \emph{external validity} is mainly concerned with the studied vulnerabilities that may not be representative enough. To reduce this risk, our study uses highly reputable bug sources (\emph{i.e.}, the Code4rena platform which has been widely used by other studies \cite{zhang2023demystifying,sun2024gptscan, xi2024pomabuster, wang2024contracttinker, liu2024using}).
As a data source, projects on Code4rena offer qualitative advantages. They provide ground truth \cite{xi2024pomabuster} and consist of vulnerabilities from real-world projects that have been confirmed by both project owners and domain experts on the platform. These vulnerabilities not only involve real assets, but also offer substantial monetary rewards to successful auditors. This indicates that our sample can represent the real threats faced by high-value contracts. Due to the lack of data sources of comparable quality, we selected only 116 target vulnerabilities from a total of 1,361 high-risk vulnerabilities. However, note that they constitute a representative sample selected according to a strictly defined and transparent process as given in Sec~\ref{SecData} and they have been made publicly available in our repository. Thus, the current distribution can still serve as a reference for real-world distributions involving substantial assets. 
Another \emph{external validity} is that the proof-of-concept checker is evaluated only on 208 active, popular GitHub projects. To mitigate this threat, we plan to conduct systematic evaluations in the future to see its more general effectiveness.

The \emph{construct validity} is mainly related with the correctness of the developed programs during our study. To alleviate this threat, we performed thorough testing to ensure the correctness of our developed programs, in particular the proof-of-concept checker. 

\section{Related Work}
\label{Related Work}
This section reviews some work closely related with this article, including empirical study on smart contract, detection tools for vulnerabilities in smart contracts, and smart contract testing.

\subsection{Empirical Study on Smart Contract}
To gain a deeper understanding of smart contract security issues, numerous empirical studies have conducted
in-depth analyses of contract vulnerabilities. 
Zhang \emph{et al.} \cite{zhang2023demystifying} systematically examine vulnerabilities in real-world smart contracts that are challenging for existing detection tools to identify and classify them into seven categories based on their root causes, with inconsistent state update vulnerabilities accounting for 18\% of the vulnerabilities reported in Code4rena. Majd \emph{et al.} \cite{soud2024fly} propose a classification framework for smart contract weaknesses from the perspectives of error sources and their impacts, offering a unique lens for understanding security risks within smart contracts. 
Additionally, Wang \emph{et al.} \cite{wang2023empirical} conduct an empirical study on historical vulnerability fixes in real-world Solidity projects, contributing to the body of knowledge on smart contract defect remediation and providing a theoretical foundation for advancing automated repair techniques. 
Chen \emph{et al.} \cite{chen2024demystifying} investigate various behavioral patterns exhibited in smart contract transactions associated with security attacks and analyze the effectiveness of invariants. Leveraging these enhanced invariants, they develop a tool capable of dynamically generating customized invariants based on historical transaction data, thereby enabling a more effective invariant-based protection mechanism.

 Beyond vulnerability analysis of smart contracts themselves, another stream of research focuses on Solidity’s language characteristics, aiming to enhance contract quality by improving the understanding of the language itself. 
Li \emph{et al.} \cite{li2023understanding} conduct the first quantitative study on event logging practices in Solidity and develop a static analysis tool based on their findings to optimize the use of logging statements. Their tool effectively identifies excessive gas consumption issues in popular GitHub projects. 
Huang \emph{et al.} \cite{huang2024revealing} investigate the misuse of libraries in smart contracts through a manual analysis of real-world cases and identify eight representative library misuse patterns covering the entire lifecycle of library usage, aiming to raise developers' awareness of library security. Bodell \emph{et al.} \cite{bodell2023proxy} conduct a large-scale empirical study on upgradeable smart contracts (USCs) to analyze their prevalence, design patterns, and security implications. Their study identifies diverse USC design patterns and highlights security risks related to upgrade mechanisms, access control, and contract management. Chen \emph{et al.} \cite{chen2021understanding} perform an empirical study on code reuse in Ethereum smart contracts to examine its prevalence and impact. Their analysis shows that code reuse is widespread, particularly in ERC20 token contracts.
Fang \emph{et al.} \cite{fang2023beyond} conduct a large-scale security analysis of modifiers in Ethereum smart contracts, uncover key security insights, and identify major usage patterns. 
 They construct a modifier dependency graph to represent all control and data flow relationships associated with modifiers, addressing the complex dependencies between modifiers, their variables, and related functions.

\subsection{Vulnerability Analysis in Smart Contracts}
As the complexity of smart contracts continues to increase, researchers have explored various methodologies to analyze and detect vulnerabilities in Ethereum smart contracts.

Li \emph{et al.} \cite{li2022smartfast} transform contract source code into an intermediate representation that facilitates vulnerability analysis. However, the transformation process may lead to the loss of critical semantic information. 
Wang \emph{et al.} \cite{wang2024efficiently} specifically target the detection of reentrancy vulnerabilities. They employ program slicing to analyze contract dependencies, collect potential vulnerability warnings, and validate them using symbolic execution.
Zhang \emph{et al.} \cite{zhang2024nyx} present Nyx, a static analyzer that detects front-running vulnerabilities in smart contracts. It combines Datalog-based preprocessing with SMT-based symbolic validation to improve efficiency and accuracy. Sun \emph{et al.} \cite{sun2024all} propose AVVERIFIER, a lightweight taint analysis framework for detecting vulnerabilities in smart contract address verification. 
 Utilizing static EVM opcode simulation, it systematically identifies and characterizes flaws in the address validation process. 
Gritti \emph{et al.} \cite{gritti2023confusum} develop JACKAL, an automated system for detecting confused contract vulnerabilities, a class of inter-contract security flaws in smart contracts. 
 By analyzing contract interactions, JACKAL identifies vulnerable call patterns and leverages a combination of static and dynamic analysis to generate exploit proofs. 
Bose \emph{et al.} \cite{bose2022sailfish} construct a storage dependency graph by analyzing program dependencies. The detection tool built upon this graph enables the automated identification of reentrancy and transaction-ordering dependencies. 
 While this tool successfully detects hazardous access patterns, it is unable to analyze global cross-contract data flows. 
Zheng \emph{et al.} \cite{zheng2022park} propose a parallel symbolic execution approach that introduces a dynamic forking algorithm based on process forking to accelerate vulnerability detection. 
Alhazami \emph{et al.} \cite{alhazami2023graph} implement a code clone detection technique without exposing contract code content. They introduce the concept of a code graph for the code clone detection problem and represent code as a graph through feature extraction. This enables the generation of a Graph Fingerprint that captures different levels of topological characteristics while preserving code privacy. 
Ghaleb \emph{et al.} \cite{ghaleb2023achecker} propose an access control vulnerability detection method that does not rely on predefined patterns or contract transaction histories. 
 Instead, they infer access control mechanisms in smart contracts through static data flow analysis and further distinguish whether unauthorized entities can gain control over the contract using symbolic analysis.
Otoni \emph{et al.} \cite{otoni2023solicitous} introduce a novel formal verification method for smart contracts. This approach generates formal verification conditions directly from the contract’s control flow graph (CFG) without requiring intermediate translation steps and models contract behavior using Constrained Horn Clauses (CHC) to achieve efficient automated verification. In addition to improving the accuracy of vulnerability detection, this method also enables the verification of contract functional correctness.

\subsection{ Smart Contract Testing}
Testing is one of the most effective methods to improve software quality in practice \cite{mttesting,mutationtesting,article,compileroptimization,6319229}. Similarly, in the process of smart contract development, testing has also been widely used to ensure the security and functionality of the contract code. Testing helps to enhance the reliability of the contract and avoid significant losses after smart contract deployment. 

Smart contracts face threats from various types of attacks, such as reentrancy attacks \cite{wang2024unity}, overflow attacks \cite{lai2020static}, timestamp dependence \cite{li2022smartfast}, and more, necessitating the coverage of various scenarios. Initially, a large amount of research has focused on using fuzz testing to detect security vulnerabilities in smart contracts \cite{wu2024we, ivanov2023security}. These studies typically define predefined detection patterns for each type of vulnerability and monitor the underlying behavior of the contract during execution. For instance, ContractFuzzer \cite{jiang2018contractfuzzer}, proposed by Jiang \emph{et al.}, detects seven known types of vulnerabilities by generating a large number of random test inputs. It also implements inter-function calls between multiple contracts, and thus is able to test the correctness of interactions with external contracts. Smartian \cite{choi2021smartian}, proposed by Choi \emph{et al.}, integrates static and dynamic data flow analysis into smart contract fuzz testing to detect security vulnerabilities in hard-to-reach branches. Solmigrator \cite{zhang2025automated}, proposed by Zhang \emph{et al.}, focuses on generating tests that are realistic and adhere to smart contract specification usage. It first extracts code snippets from the actual use of on-chain contracts, then migrates them to newly developed smart contracts with similar functionality for testing. 

However, fuzz testing itself has inherent instability and limitations \cite{li2018fuzzing, manes2019art}, which may lead to poor repeatability of test results and difficulty in covering all program paths, thus affecting vulnerability discovery rates and accuracy. To address this, many research efforts have utilized symbolic execution \cite{CUTE,EXE,DART} to test smart contracts. Mossberg \emph{et al.} propose Manticore \cite{mossberg2019manticore}, which uses constraint-solving to systematically explore the program state space. In particular, Manticore provids a flexible dynamic symbolic execution framework that supports both traditional and external execution environment extensions. So \emph{et al.} propose SMARTEST \cite{so2021smartest}, which focuses on the state of smart contracts. More specifically, SMARTEST combines symbolic execution with a language model of vulnerable transaction sequences, effectively searching for attack-prone transaction sequences in smart contracts. Park \cite{zheng2022park}, proposed by Zheng \emph{et al.}, uses multiple processes during symbolic execution and utilizes multiple CPU cores to improve efficiency. To address the performance loss of SMT in parallelization, they additionally propose an adaptive process limit and adjustment algorithm, as well as a shared memory-based global variable reconstruction method to collect and rebuild global variables from different processes. However, symbolic execution still faces the state explosion problem, especially when dealing with large programs or complex inputs, often leading to an analysis process that cannot be completed within a reasonable time frame \cite{baldoni2018survey, cadar2011symbolic}.

\subsection{Benchmarks for Smart Contract Testing and Analysis}
Some research efforts focus on building benchmarks to test and analyze smart contracts.
Durieux \emph{et al.} \cite{durieux2020empirical} construct datasets through web scraping, which include annotated vulnerable smart contracts and smart contracts deployed on the Ethereum blockchain. They also create a new scalable execution framework named SmartBugs, and conduct an empirical evaluation of nine state-of-the-art automated analysis tools. 
Ghaleb \emph{et al.} \cite{ghaleb2020effective} propose an automated and systematic approach, SolidiFI, for evaluating static analysis tools for smart contracts. To conduct the evaluation, this approach injects code defects into smart contracts to introduce targeted security vulnerabilities.
To understand the security level of smart contracts, Hwang \emph{et al.} \cite{hwang2020gap} collect and publicly release over 50,000 real smart contracts deployed on Ethereum. They further identify and analyze the vulnerabilities that have been patched by the Solidity compiler to gain deeper insights into the security patches of the Solidity compiler.
Ren \emph{et al.} \cite{ren2021empirical} propose a unified evaluation standard to eliminate biases in the evaluation process. They collect 46,186 available smart contracts from four influential organizations. This comprehensive dataset, which is open to the public, includes a wide range of code attributes, vulnerability trends, and use cases.
Li \emph{et al.} \cite{li2024static} propose a new fine-grained taxonomy that includes 45 distinct types of smart contract vulnerabilities. Based on this taxonomy, they develop a comprehensive benchmark that covers various code features, vulnerability patterns, and application scenarios.

\section{Conclusion}
\label{Conclusion}
Inconsistent state update bugs are common in practice and can result in vulnerability attacks. To the best of our knowledge, this paper presents the first systematic empirical study of inconsistent state update vulnerabilities in the wild. In particular, we systematically investigate 116 inconsistent state update vulnerabilities in 352 real-world smart contract projects, summarizing their root causes, fix strategies, and  exploitation methods. By doing so, we present 11 original and important findings with actionable implications. These findings and implications are beneficial to developers, researchers, tool builders, and language or library designers. To illustrate the potential benefits of our research, we also develop a proof-of-concept checker based on one of our findings. The checker effectively detects issues in 64 active, popular GitHub projects, and 19 project owners have confirmed the detected issues at the time of writing. 

Looking ahead, our primary goal is to develop more advanced tools based on our research findings. In particular, we aim at enabling them to comprehend complex semantic relationships among variables and thereby more comprehensively detect inconsistent state update vulnerabilities in the wild. Second, the taxonomy of root causes and fix strategies we have constructed can be used to build intelligent repair systems that provide empirically grounded fix suggestions or even automatically generate patches for detected vulnerabilities. Third, future work could extend the study of Ethereum smart contracts in this paper to other blockchain platforms, and investigate how different programming paradigms and virtual machine architectures affect the prevalence, manifestations, and root causes of inconsistent state update vulnerabilities. Finally, conducting large-scale longitudinal studies to monitor how the inconsistent state update vulnerabilities evolve with new language features and development practices will help the community proactively address emerging threats.

\section{Data Availability}
Our replication package (including code, dataset, \emph{etc.}) is publicly available at \url{https://github.com/logicseek/StateStudy}.

\section*{Acknowledgments}
\noindent
We deeply appreciate the reviewers for their insightful comments. This work was supported by National Natural Science Foundation of China (Grant No. 62102233), Shandong Province Overseas Outstanding Youth Fund (Grant No. 2022HWYQ-043), Joint Key Funds of National Natural Science Foundation of China (Grant No. U24A20244), and Qilu Young Scholar Program of Shandong University.

\bibliographystyle{IEEEtran}
\bibliography{sample-base}

\end{document}